%% file: B2G-23-006_temp.tex
\pdfoutput=1
\documentclass[11pt,twoside,a4paper,cmspaper,final,collab]{cms-tdr}
\def\svnVersion{d12d394}\def\svnDate{2025/05/22}\def\cmsCernNoTag{CERN-EP-2024-301}\def\cmsCernDate{\today}\def\cmsMessage{Published in Physics Letters B as \href{http://dx.doi.org/10.1016/j.physletb.2025.139568}{\doi{10.1016/j.physletb.2025.139568}.}}
\begin{document}\cmsNoteHeader{B2G-23-006}

\newcommand{\SD}{\unit{SD}}
\newcommand{\CP}{{\itshape CP}\xspace}
\newcommand{\pp}{\ensuremath{\Pp\Pp}\xspace}
\newcommand{\AZH}{\ensuremath{\PSA\to\PZ\PSH}\xspace}
\newcommand{\ppAZH}{\ensuremath{\pp\to\PSA\to\PZ\PSH}\xspace}
\newcommand{\Httbar}{\ensuremath{\PSH\to\ttbar}\xspace}
\newcommand{\AZHttZ}{\ensuremath{\PSA\to\PZ\PSH\to\PZ\ttbar}\xspace}
\newcommand{\xsbr}{\ensuremath{\sigma(\pp\to\PSA)\mathcal{B}(\PSA\to\PZ\PSH)\mathcal{B}(\Httbar)}\xspace}
\newcommand{\ttZ}{\ensuremath{\ttbar\PZ}\xspace}
\newcommand{\ttW}{\ensuremath{\ttbar\PW}\xspace}
\newcommand{\ZG}{\ensuremath{\PZ/\PGg^{\ast}}\xspace}
\newcommand{\ZGjets}{\ensuremath{\PZ/\PGg^{\ast}\text{+jets}}\xspace}
\newcommand{\ZGHF}{\ensuremath{\PZ/\PGg^{\ast}\text{+HF}}\xspace}
\newcommand{\ZGLF}{\ensuremath{\PZ/\PGg^{\ast}\text{+LF}}\xspace}
\newcommand{\singlet}{\text{single \PQt}\xspace}
\newcommand{\VV}{\ensuremath{\PV\PV}\xspace}
\newcommand{\mA}{\ensuremath{m_{\PSA}}\xspace}
\newcommand{\mH}{\ensuremath{m_{\PSH}}\xspace}
\newcommand{\mttZ}{\ensuremath{m_{\ttZ}}\xspace}
\newcommand{\mtt}{\ensuremath{m_{\ttbar}}\xspace}
\newcommand{\DeltaM}{\ensuremath{\Delta m}\xspace}
\newcommand{\ptZ}{\ensuremath{\pt^{\PZ}}\xspace}
\newcommand{\mZ}{\ensuremath{m_{\PZ}}\xspace}
\newcommand{\mll}{\ensuremath{m_{\Pell\Pell}}\xspace}
\newcommand{\finalObs}{\ensuremath{\ptZ\times\DeltaM}\xspace}
\newcommand{\GammaA}{\ensuremath{\Gamma_{\PSA}}\xspace}
\newcommand{\mumu}{\ensuremath{\PGmp\PGmm}\xspace}
\newcommand{\ee}{\ensuremath{\Pep\Pem}\xspace}

\cmsNoteHeader{B2G-23-006}
\title{Search for heavy neutral Higgs bosons \texorpdfstring{\PSA}{A} and \texorpdfstring{\PSH}{H} in the \texorpdfstring{\ttZ}{ttZ} channel in proton-proton collisions at \texorpdfstring{13\TeV}{13 TeV}}
\author{The CMS Collaboration}
\date{\today}

\abstract{
  A direct search for new heavy neutral Higgs bosons \PSA and \PSH in the
  \ttZ channel is presented, targeting the process \ppAZH with \Httbar. For the first time, the channel with decays of the \PZ boson to muons or electrons in association with all-hadronic decays of the \ttbar system is targeted. The analysis uses proton-proton collision data collected at the CERN LHC with the CMS experiment at $\sqrt{s}=13\TeV$, which correspond to an integrated luminosity of 138\fbinv. No signal is observed. Upper limits on the product of the cross section and branching fractions are derived for narrow resonances \PSA and \PSH with masses up to 2100 and 2000\GeV, respectively, assuming \PSA boson production through gluon fusion. The results are also interpreted within two-Higgs-doublet models, where \PSA and \PSH are \CP-odd and \CP-even states, respectively, complementing and substantially extending the reach of previous searches.
}

\hypersetup{%
pdfauthor={CMS Collaboration},%
pdftitle={Search for heavy neutral Higgs bosons A and H in the ttZ channel},%
pdfsubject={Search for heavy neutral Higgs bosons A and H in the ttZ channel},%
pdfkeywords={CMS, 2HDM}}

\maketitle

\section{Introduction}\label{sec:introduction}
The measured properties of the Higgs boson discovered~\cite{ATLAS:2012yve,CMS:2012qbp,CMS:2013btf} by the ATLAS and CMS Collaborations at the CERN LHC are consistent with the expectations of the standard model (SM) within the current experimental precision~\cite{ATLAS:2022vkf,CMS:2022dwd}.
However, the particle might be part of an extended Higgs sector, as suggested by several models of physics beyond the SM (BSM), such as supersymmetry~\cite{Djouadi:2005gj}, models with an axion~\cite{Kim:2008hd} or dark matter~\cite{Bauer:2017ota}, as well as BSM mechanisms of baryogenesis~\cite{Biekotter:2022kgf}.

A common class of extended Higgs models is two-Higgs-doublet models (2HDMs)~\cite{Lee:1973iz,Branco:2011iw}, containing one additional Higgs doublet, resulting in the presence of three neutral Higgs bosons, two of which (\PSh and \PSH) are even under charge-parity (\CP) transformations and one (\PSA) is \CP-odd, as well as two charged Higgs bosons (\PSHpm).
Such models are defined by the following parameters when assuming \CP conservation:
the masses of the five Higgs bosons $m_{\PSh}$, \mH, \mA, and $m_{\PSHpm}$,
the ratio of the vacuum expectation values of the two doublets \tanb,
the mixing angle between the two \CP-even Higgs bosons $\alpha$,
and the soft-breaking term $m_{12}^{2}$.
Depending on the coupling of the two doublets to the right-handed quarks and the charged leptons, different types of 2HDM are possible.
For example, in type-I models the quarks and charged leptons couple to a single Higgs doublet, while in type-II models the down-type quarks and the charged leptons couple to a different Higgs doublet from that coupling to the up-type quarks.
When $\cos(\beta-\alpha)\to0$, referred to as the ``alignment limit'', the properties of the \PSh boson are expected to be identical to those of a SM Higgs boson with the same mass.
In this Letter, \PSh is identified with the observed Higgs boson, which has a mass of approximately 125\GeV.

For sufficiently large mass splittings, $\mA>\mH+\mZ$, the decay \AZH becomes kinematically possible and is dominant in a wide range of the 2HDM parameter space~\cite{Haisch:2018djm}.
Searches for \AZH have been performed in final states targeting \PSH boson decays to \bbbar, $\PW\PW$, and $\PGtp\PGtm$~\cite{ATLAS:2020gxx,CMS:2019ogx,CMS:2016xnc}.
Their sensitivity is typically limited to masses of \mH up to approximately 400\GeV, beyond which the \Httbar decay becomes kinematically possible and is dominant, in particular for small and moderate values of \tanb.
In models of electroweak baryogenesis that attempt to explain the observed matter-antimatter asymmetry in the universe, a first-order electroweak phase transition becomes possible at these values of masses and \tanb~\cite{Dorsch:2013wja,Basler:2016obg,Biekotter:2022kgf}. However, this region of parameter space has not been probed extensively in experimental analyses.
Searches for $\PSA/\PSH\to\ttbar$ decays provide some sensitivity to the relevant parameter space~\cite{ATLAS:2024vxm,CMS:2019pzc}, and indirect constraints are obtained, for example, from $\PQt\PAQt\PQt\PAQt$ cross section measurements~\cite{ATLAS:2018alq,CMS:2019rvj}.
Recently, the ATLAS Collaboration published a direct search for \AZH with \Httbar in final states corresponding to single-lepton decays of the \ttbar system, excluding at 95\% confidence level (\CL) regions up to approximately $\mA=1\TeV$ for $\tanb=1$ of the type-I 2HDM parameter space. They report an excess in the region around $(\mA,\mH)=(650,450)\GeV$ with a local significance of
2.85 standard deviations (SD)~\cite{ATLAS:2023zkt}.

In this Letter, a direct search for heavy neutral Higgs bosons \PSA and \PSH is presented, targeting the process \AZH with \Httbar.
The final state with decays of the \PZ boson to muons or electrons in association with all-hadronic decays of the \ttbar system is targeted, which has not been studied before.
The search is performed with proton-proton (\pp) collision data recorded at $\sqrt{s}=13\TeV$ at the LHC during 2016--2018, corresponding to a total integrated luminosity of 138\fbinv~\cite{CMS-LUM-17-003,CMS-PAS-LUM-17-004,CMS-PAS-LUM-18-002}.
Events are selected by requiring the presence of two same-flavour and oppositely charged leptons ($\ell=\Pgm, \Pe$) that are compatible with originating from a \PZ boson decay, as well as at least five jets.
Two sensitive observables are reconstructed, following Ref.~\cite{Haisch:2018djm}:
\DeltaM, characterizing the mass difference of the \PSA and the \PSH boson
candidates, and \ptZ, characterizing the transverse momentum of the \PZ boson candidate.
The two-dimensional distribution of \DeltaM and \ptZ, reduced into a one-dimensional histogram, is used as the final discriminating observable in the search.
Upper limits on the product of the cross section and branching fractions \xsbr are derived as a function of \mA and \mH, assuming gluon fusion production, and narrow \CP-odd and \CP-even resonances \PSA and \PSH, respectively.
The results are used to constrain the 2HDM parameter space.
Tabulated results are provided in the HEPData record for this analysis~\cite{HEPData}.

\section{The CMS detector and event reconstruction}\label{sec:reconstruction}
The CMS apparatus~\cite{CMS:2008xjf,CMS:2023gfb} is a multipurpose, nearly hermetic detector, designed to trigger on~\cite{CMS:2020cmk,CMS:2016ngn,CMS:2024aqx} and identify muons, electrons, photons, and (charged and neutral) hadrons~\cite{CMS:2020uim,CMS:2018rym,CMS:2014pgm}. A global ``particle-flow'' (PF) algorithm~\cite{CMS:2017yfk} aims to reconstruct all individual particles in an event, combining information provided by the all-silicon inner tracker and by the crystal electromagnetic (ECAL) and brass-scintillator hadron calorimeters, operating inside a 3.8\unit{T} superconducting solenoid, with data from the gas-ionization muon detectors embedded in the flux-return yoke outside the solenoid. The reconstructed particles are used to build \PGt leptons, jets, and missing transverse momentum~\cite{CMS:2018jrd,CMS:2016lmd,CMS:2019ctu}.

Events of interest are selected using a two-tiered trigger system.
The first level, composed of custom hardware processors, uses information from the calorimeters and muon detectors to select events at a rate of around 100\unit{kHz} within a fixed latency of about 4\mus~\cite{CMS:2020cmk}.
The second level, known as the high-level trigger, consists of a farm of processors running a version of the full event reconstruction software optimized for fast processing, and reduces the event rate to around 1\unit{kHz} before data storage~\cite{CMS:2016ngn}.

The primary vertex is taken to be the vertex corresponding to the hardest scattering in the event, evaluated using tracking information alone, as described in Section 9.4.1 of Ref.~\cite{CMS-TDR-15-02}.

The reconstruction of muons relies on a combination of measurements in the tracker and in the muon detectors~\cite{CMS:2018rym}.
The muons are identified based on the quality of the combined track fit and on the number of hits in the different tracking detectors, with an efficiency of about 95\%.
The transverse momentum (\pt) resolution in the barrel is better than 1\% (7\%) for muons with \pt up to 100\GeV (1\TeV)~\cite{CMS:2018rym}. 
Electrons are reconstructed by combining the momentum measurement in the tracker with the energy measurement of the corresponding cluster in the ECAL and with the energy sum of all bremsstrahlung photons, obtained from the ECAL, spatially compatible with originating from the electron track~\cite{CMS:2020uim,CMS-DP-2020-021}.
The electrons are identified using criteria on the cluster shape in the ECAL, the track quality, and the compatibility between the tracker and ECAL measurements, corresponding to an identification efficiency of approximately 80\%, including the isolation requirements described below.
The momentum resolution for electrons with $\pt \approx 45\GeV$ from $\PZ \to \Pe \Pe$ decays ranges from 1.6 to 5\%~\cite{CMS:2020uim,CMS-DP-2020-021}.
Both muons and electrons are required to lie within the acceptance of the tracker, covering pseudorapidity ($\eta$) values of $\abs{\eta}<2.4$, and to have $\pt>35$ and 20\GeV for the \pt-leading and subleading leptons, respectively.
The \pt thresholds were chosen to maximize the expected sensitivity of the search to a potential signal, evaluated using simulation.
Electrons reconstructed in the transition region $1.44<\abs{\eta}<1.56$ between the barrel and the endcap calorimeters are discarded.
Isolation requirements are imposed on the leptons with the aim to suppress leptons originating from decays of heavy hadrons inside jets.
The isolation is based on the scalar \pt sum of all PF objects in a cone of radius 0.3 in the $\eta$-$\phi$ plane around the track direction of the lepton, where $\phi$ denotes the azimuthal angle, and it is corrected by removing contributions arising from additional \pp interactions in the same bunch crossing (pileup)~\cite{CMS:2020ebo}.

Hadronic jets are clustered from the PF objects using the anti-\kt algorithm~\cite{Cacciari:2008gp,Cacciari:2011ma} with a distance parameter of 0.4, omitting charged particles matched to pileup vertices.
The jet energy is corrected for the neutral-hadron contribution expected from pileup interactions~\cite{Cacciari_2008}.
Further energy corrections depending on the jet \pt and $\eta$ are applied, which are derived in simulation such that the average measured energy of the jets becomes identical to that of the particle-level jets.
Residual differences between the jet energy scale (JES) in data and in simulation are measured using the momentum balance in dijet, $\gamma$+jets, \PZ{}+jets, and multijet events, and appropriate corrections are applied.
The jet energy resolution (JER) typically amounts to 15--20\% at 30\GeV, 10\% at 100\GeV, and 5\% at 1\TeV.
Jets overlapping within a cone of radius 0.4 in the $\eta$-$\phi$ plane with a muon or an electron passing the criteria described above are discarded.
Further selection criteria are applied to each jet to remove jets potentially dominated by instrumental effects or reconstruction failures, as well as jets arising from pileup interactions~\cite{CMS:2020ebo}.
Only jets within the tracker acceptance of $\abs{\eta}<2.4$ and with $\pt>30\GeV$ are considered in the analysis.
Jets arising from the hadronization of \PQb quarks (\PQb jets) are
identified using the \textsc{DeepJet} \PQb tagging algorithm~\cite{BTV-16-002,Bols:2020bkb,CMS-DP-2023-005}.
Jets are considered as \PQb tagged if they fulfill a requirement on the
\textsc{DeepJet} discriminant that corresponds to a \PQb tagging efficiency of
75--80\% and a 1.5--2\% (15--17\%) misidentification rate for light-flavour quark
and gluon (\PQc quark) induced jets.

\section{Simulated signal and background event samples}\label{sec:samples}
Several Monte Carlo (MC) event generators, interfaced with a detailed detector simulation based on \GEANTfour~v.\ 9.4~\cite{Agostinelli:2002hh}, are used to model the signal and background events.
The proton structure is described by the parton distribution function (PDF) set {NNPDF3.1}~\cite{Ball:2017nwa}.
Parton showering (PS) and hadronization are simulated with \PYTHIA~v.\ 8.240~\cite{Sjostrand:2014zea}, where the parameters for the underlying event description correspond to the CP5 tune~\cite{Sirunyan:2019dfx}.

The $\pp\to\AZHttZ$ signal is simulated in the gluon fusion production mode at leading order (LO) precision in QCD using \MGvATNLO~v.\ 2.6.5~\cite{Alwall:2014hca,Degrande:2014vpa,Degrande:2011ua}.
Events are generated for different mass configurations of the \PSA and \PSH bosons, modelled as narrow resonances with relative decay widths $\Gamma_{\PSA/\PSH}/m_{\PSA/\PSH}=3$\%, which is negligible compared to the experimental resolution.
Events are additionally generated assuming larger widths for some mass configurations, in order to estimate the effect of the total decay width in the 2HDM interpretation.
Interference effects between the signal and SM background from \ttbar production in
association with a \PZ boson (\ttZ), which is a minor background in the analysis, are small.
They modify the rate and the shape of the \DeltaM and \ptZ distributions by a few percent at most~\cite{Haisch:2018djm}, and are therefore neglected.
Other \PSA boson production modes in the 2HDM, for which \PQt and \PQb quark associated production have the largest cross sections, contribute less than 2\% of the cross section for $\tanb<2$, which is the parameter region relevant for this analysis, and are therefore neglected.

The final state targeted by this search is dominated by background processes from \ttbar production and production of \ZG in association with jets referred to as \ZGjets.
The \ttbar background process is simulated at next-to-LO (NLO) precision using \POWHEG~v.\ 2~\cite{Nason:2004rx,Frixione:2007vw,Alioli:2010xd,Jezo:2015aia,Frixione:2007nw}, and the inclusive cross section is normalized to a calculation at next-to-NLO (NNLO) in QCD including resummation of next-to-next-to-leading logarithmic (NNLL) soft-gluon terms~\cite{bib:ttbarxs:1,bib:ttbarxs:2,bib:ttbarxs:3,bib:ttbarxs:4,bib:ttbarxs:5,bib:ttbarxs:6,bib:ttbarxs:7}.
The top quark \pt spectrum in simulation has been found to be harder than the one observed in data~\cite{CMS:2018htd,CMS:2024ybg}, and therefore, the simulated events are corrected as a function of the top quark and antiquark \pt at the parton level, such that the corrected simulation matches the data.

The \ZGjets background processes are simulated at LO precision with up to four partons at the matrix element (ME) level with the \MGvATNLO generator, where the merging of the jets from ME calculations and the PS is done using the MLM~\cite{Alwall:2007fs} prescription.
The \ZGjets events are separated into two mutually exclusive processes, based on the flavour of the jets at the particle-level, where these jets fulfill the acceptance requirements of $\pt>20\GeV$ and of $\abs{\eta}<2.4$.
Events with at least one particle-level jet that contains one or more \PQb- or \PQc-flavoured hadrons are classified as \ZG + heavy-flavour jets (\ZGHF) events, and all other events are classified as \ZG + light-flavour and gluon jets (\ZGLF) events.
Both \ZGHF and \ZGLF events are corrected depending on the vector boson \pt to calculations at NLO precision in electroweak and QCD corrections~\cite{Lindert:2017olm}.

Minor background contributions arise from \ttZ production and from \ttbar production in
association with a \PW boson (\ttW), single top quark
production in the $t$- and $s$ channels and $\PQt\PW$ production, collectively referred to as \singlet, as well as from $\PW\PW$,
$\PW\PZ$, and $\PZ\PZ$ production, collectively referred to as \VV.
The \ttZ and \ttW processes are simulated at NLO precision with \MGvATNLO using the \textsc{FxFx}~\cite{Frederix:2012ps} prescription for the ME-PS jet matching, and their inclusive cross sections are scaled to take into account NLO+NNLL QCD and NLO electroweak corrections~\cite{Kulesza:2018tqz}.
The \singlet processes are simulated at NLO precision, in the $t$ and $\PQt\PW$ channels with \POWHEG~\cite{powheg:singlet:st,powheg:singlet:tW}, and in the $s$ channel with \MGvATNLO.
The inclusive cross sections of the $t$ and $s$ channel processes are normalized to the predictions at NLO precision~\cite{bib:singletxs:t:1,bib:singletxs:t:2} and the $\PQt\PW$ channel to calculations at approximate NNLO precision~\cite{bib:singletxs:tw:1}.
Diboson production is simulated at LO precision using \PYTHIA, and the inclusive cross sections are normalized to NNLO precision~\cite{Gehrmann:2014fva} for the $\PW\PW$ process and to NLO precision~\cite{bib:mcfm:diboson} for the other diboson processes.
Contributions from triboson processes, such as $\PW\PW\PZ$ production, are below the per mille level of the total background and are therefore neglected.

For comparison with the observed distributions, the events in the simulated samples are normalized to the same integrated luminosity as the data sample, according to their predicted cross sections.
Pileup effects are modelled by adding simulated minimum bias events to all simulated events with a frequency distribution adjusted to match that observed in the data.

\section{Event selection and analysis strategy}\label{sec:strategy}
The data are selected using a combination of triggers.
The triggers require the presence of either one or two muon candidates for the \mumu events, and likewise either one or two electron candidates for the \ee events.
In addition, to maximize the signal selection efficiency for \ee events, the analysis uses triggers requiring the presence of one or two photon candidates, which at the trigger level can originate from electrons.
The selection efficiency with respect to the offline reconstruction achieved for signal events after the event
selection described below is above 99\% in the \mumu channel in
the entire $(\mA,\mH)$ mass plane and above 99 or 97\% in the \ee
channel for \mH above or below 300\GeV, respectively.
The single-object triggers contribute about 6--8\% to the signal selection efficiency.

\begin{figure}[!htp]
  \centering
  \includegraphics[width=0.5\textwidth]{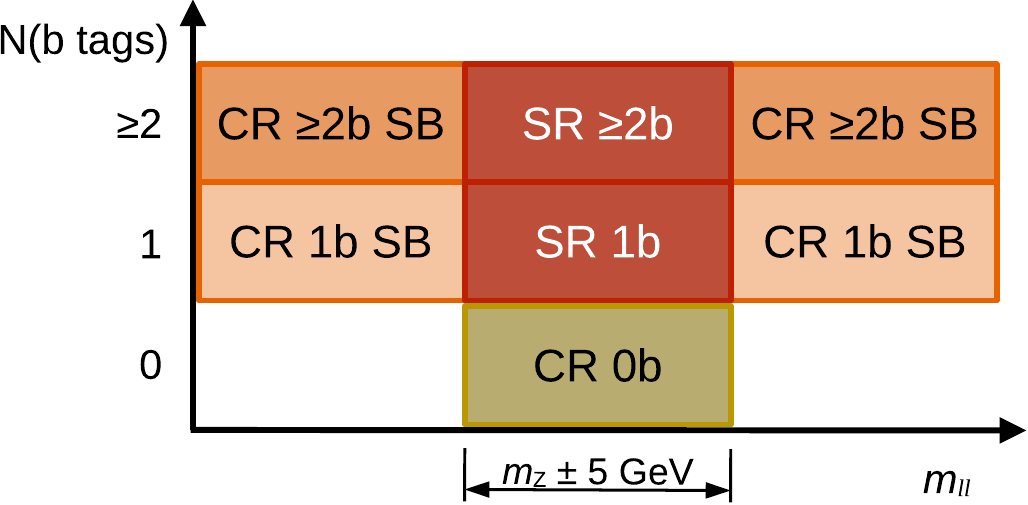}
  \caption{
    Definition of the event categories.
    The five partitions shown apply to each of the classes of jet multiplicity (5j and $\geq$6j), resulting in 10 categories per lepton flavour channel (\mumu or \ee). 
    A value of 91\GeV is used for the \PZ boson mass \mZ in the definition of the \PZ window.
  }
  \label{fig:categorization}
\end{figure}

A further offline selection requires events to have exactly two
muons or two electrons of opposite charge, and at least five jets.
The invariant mass of the selected lepton pair, \mll, is required to
be larger than 30\GeV, and \ptZ to be larger
than 15\GeV to suppress events from heavy-flavour resonance
decays and low-mass Drell--Yan processes.
The selected events are then categorized based on the lepton flavour
into two mutually exclusive channels, \mumu and \ee.
In each channel, events are categorized into events with five jets
(5j) and into events with six or more jets ($\geq$6j), and, in each jet
multiplicity category, further into events with two or more
\PQb-tagged jets ($\geq$2b), one \PQb-tagged jet (1b), and zero \PQb-tagged
jets (0b).
In each of these six categories in jet and \PQb-tagged jet multiplicity
per channel, events are finally separated into one region with $\abs{\mll-91\GeV}<5\GeV$ (\PZ window), \ie close to the \PZ boson mass, and
in one region with $\abs{\mll-91\GeV}>5\GeV$ (sideband, SB).
In the 0b region, only the events with \mll inside the \PZ window
region are used further in the analysis.
This categorization scheme is illustrated in
Fig.~\ref{fig:categorization} and leads to ten categories in total per
lepton-flavour channel.

The $\geq$2b and 1b categories inside the \PZ window are the signal regions (SRs). 
The overall signal selection efficiency achieved in the four SRs together ranges between 3 and 13\%,
depending on \mA and \mH.
The 5j SRs mainly recover signal events where one jet falls outside the experimental acceptance and, to a lesser extent, signal events where two nearby jets are merged into one single jet, and contribute up to 50\% to the total signal selection efficiency at low \mA.
Events with fewer than five jets constitute ${<}4$\% of the signal and therefore are not considered in the analysis.

The normalizations of the overall event yields of the dominant \ttbar and \ZGjets background processes are free parameters of the fit to data described in Section~\ref{sec:results}, providing flexibility to adjust the predicted yields to those observed in data.
The normalizations of the minor background processes are taken from the calculations discussed in Section~\ref{sec:samples}.
The SB and the 0b categories are control regions (CRs) that are enriched in \ttbar and \ZGjets events, respectively.
Inclusion of the CRs in the fit increases its sensitivity to the normalization parameters and helps to reduce the systematic uncertainties discussed in Section~\ref{sec:systematics}.

In each selected event, \mH is reconstructed as the invariant mass of the \ttbar system, \mtt, computed as the invariant mass of the reconstructed jets.
If there are five or six reconstructed jets in the event, these are used to compute \mtt.
If there are more than six reconstructed jets, \mtt is computed as the invariant mass of the six jets that provide the best compatibility with the $\ttbar\to\PQb\PAQb\PQq\PAQq\PQq\PAQq$ decay hypothesis.
The six jets are selected by minimizing a $\chi^{2}$ function, which quantifies the compatibility of the invariant masses of the different two- and three-jet pairings with those of the known masses of the \PW boson and the \PQt quark from which the jets originate under the $\ttbar\to\PQb\PAQb\PQq\PAQq\PQq\PAQq$ decay hypothesis.
The $\chi^{2}$ function takes into account the experimental resolutions in the two- and three-jet invariant masses of 17\GeV and 28\GeV, respectively.
The two leptons from the \PZ boson decay are then included to calculate \mttZ, the invariant mass of the \ttZ system.

Finally, two sensitive observables are constructed to discriminate the signal from the background~\cite{Haisch:2018djm}.
These exploit kinematic properties of the predominantly on-shell
production of the \PSA and \PSH bosons in the \AZHttZ process.
The distributions in \mttZ and \mtt feature Breit--Wigner peaks at
approximately \mA and \mH, respectively, and their difference
{$\DeltaM = \mttZ - \mtt \approx \mA - \mH$}
is used as an observable of interest.
Since the four-momenta of the \PSH and \PZ bosons are
determined by \mA, the \ptZ spectrum has a characteristic shape with a
kinematic edge~\cite{Haisch:2018djm} and is used as the second observable of interest, reconstructed as the
\pt of the dilepton system.
The variables \DeltaM and \ptZ provide a better experimental resolution than \mttZ and \mtt through a reduction of effects from jet mismeasurements.
The reconstructed \ptZ depends only on the excellent lepton momentum
resolution and deviations in the reconstructed values of \mttZ and
\mtt from the true values caused by mismeasurements of jet momenta are
correlated and are reduced when using the mass difference \DeltaM.

\begin{figure}[!htp]
  \centering
  \includegraphics[width=0.45\textwidth]{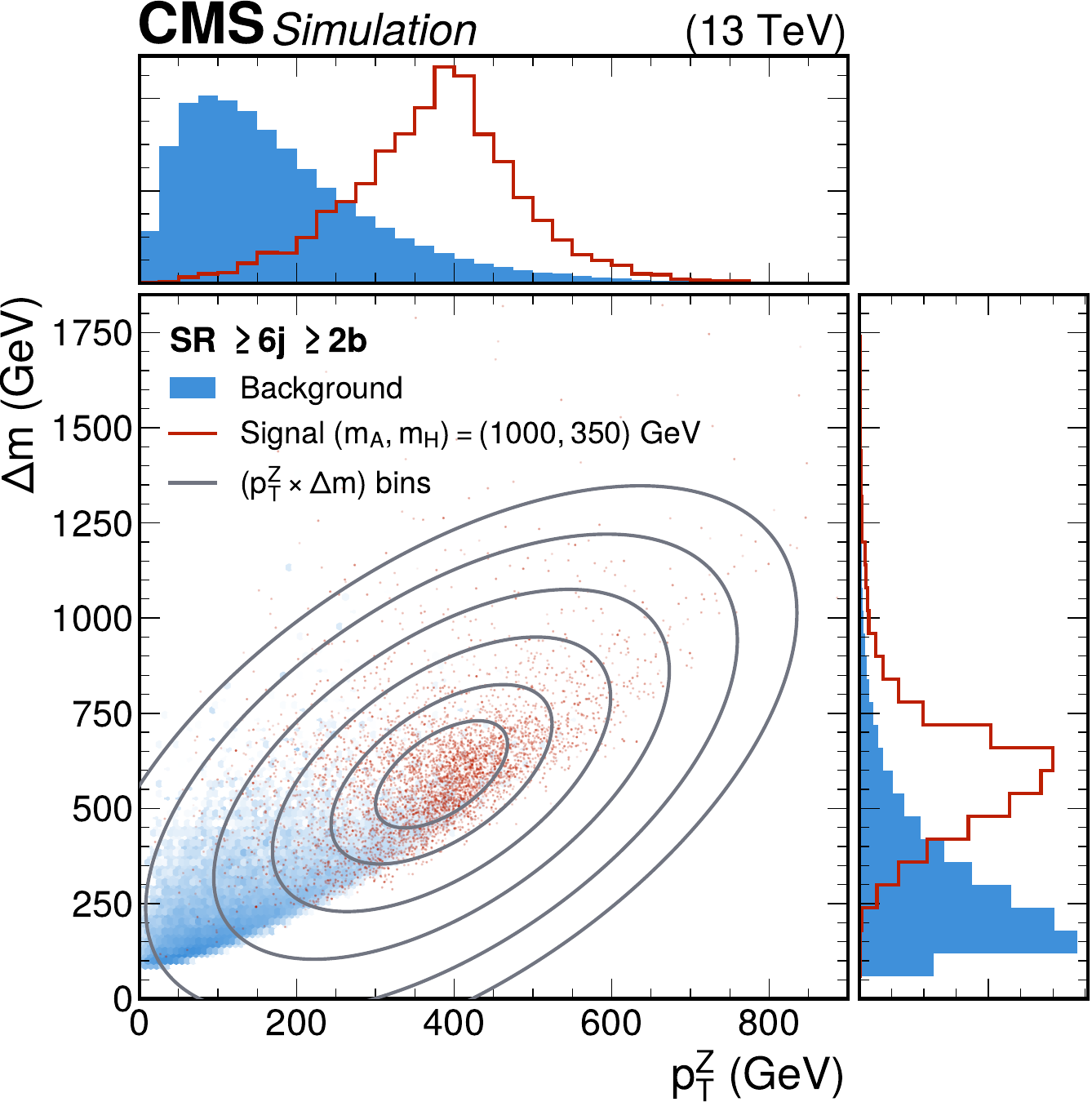}
  \caption{
    Expected $(\DeltaM,\ptZ)$ distribution for signal events with
    $\mA=1000\GeV$ and $\mH=350\GeV$ (red) and sum of backgrounds (blue) in the \mbox{SR $\geq$6j $\geq$2b} category, and boundaries of the elliptical bins (solid lines) used to construct the final observable.
  }
  \label{fig:ellipses}
\end{figure}

\begin{figure*}[!htp]
  \centering
  \includegraphics[width=0.7\textwidth]{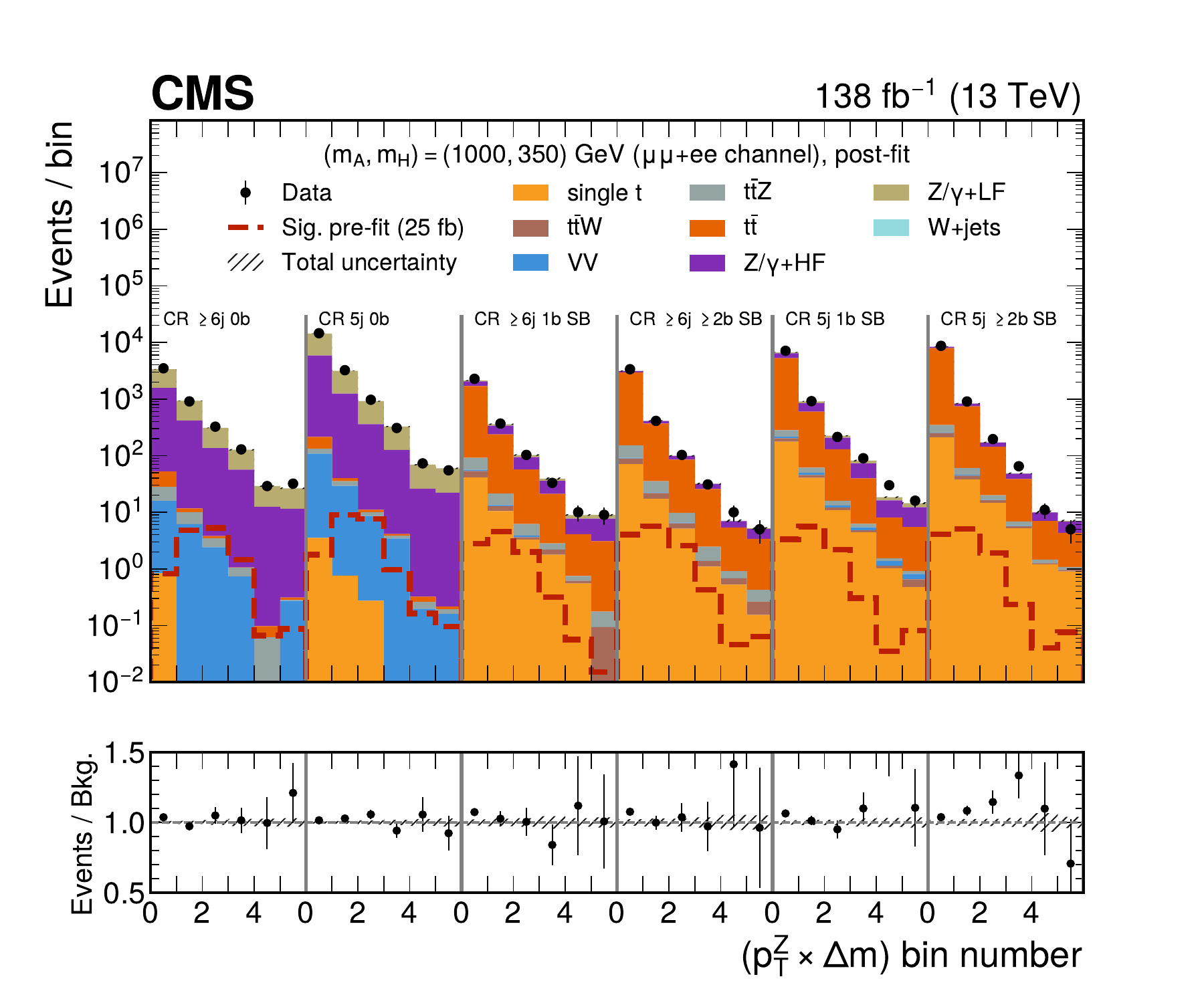}
  \includegraphics[width=0.7\textwidth]{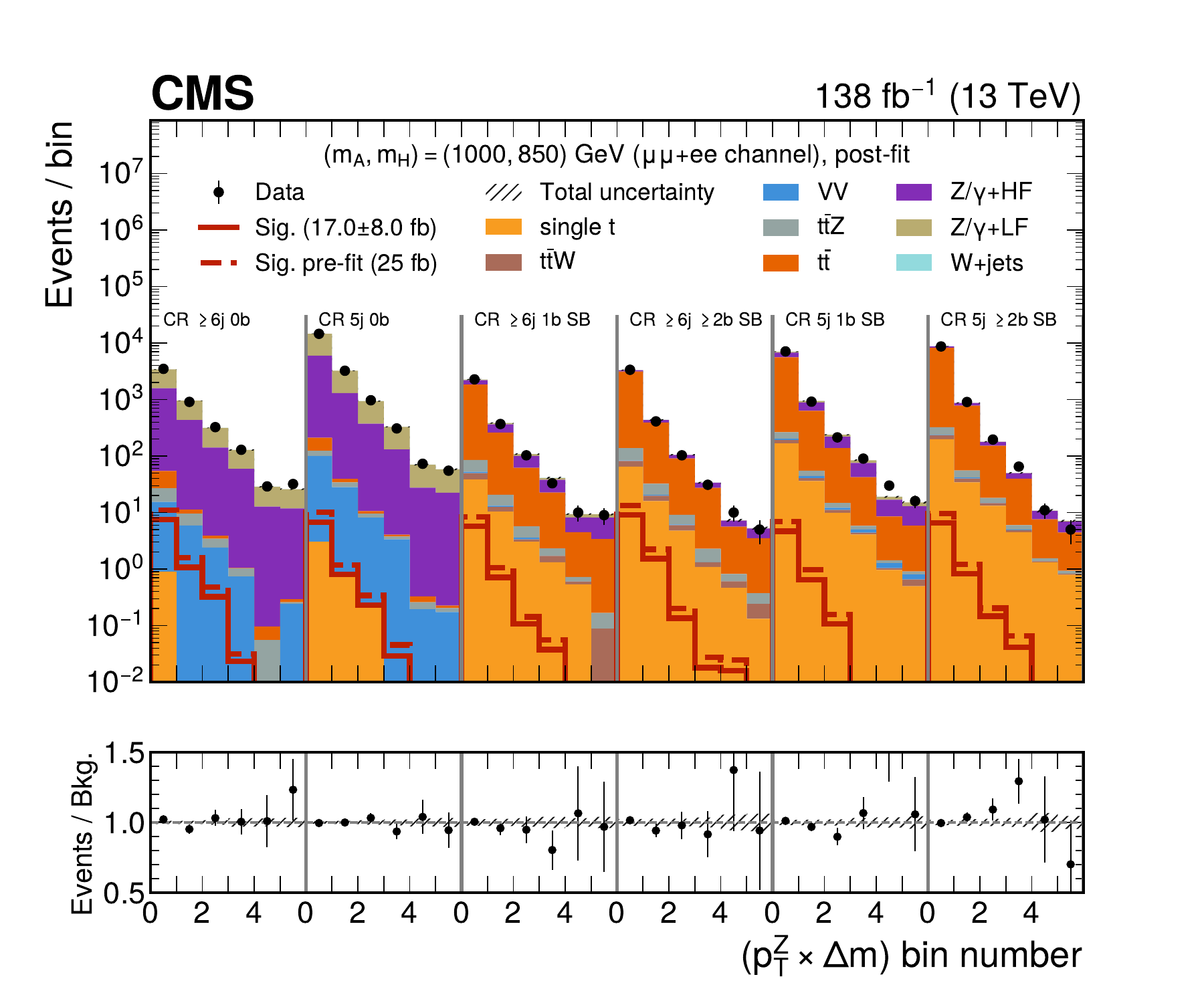}
  \caption{
    Distributions of events in the \finalObs bins in the
    CRs after the fit to data described in Section~\ref{sec:results}
    with a signal hypothesis of $\mA=1000\GeV$ and $\mH=350$ (upper) and
    $\mH=850\GeV$ (lower).
    The signal (solid red line) and background (coloured histograms)
    distributions are shown with their best fit normalizations from
    the simultaneous fit to the data (``post-fit''), where the sum
    of the \mumu and \ee channels is shown for illustration purposes.
    In the upper plot, the post-fit signal cross section is found to be $0.0\pm0.1\unit{fb}$.
    Therefore, the corresponding histogram is not displayed.
    The signal is also shown for a normalization to 25\unit{fb} (``pre-fit'') in the dashed red line.
    The lower panels show the ratio of the observed data to the background events (black points with error bars) and the ratio of the total number of expected signal-plus-background events to the background events (red lines).
    In both panels, the hatched area represents the total uncertainty.
  }
  \label{fig:CR:2D}
\end{figure*}

The two-dimensional distribution in \DeltaM and \ptZ, reduced into a one-dimensional histogram, is used as the final observable to search for the signal. In order to capture the relevant features of the signal distributions with high statistical precision, the information is reduced to a one-dimensional distribution \finalObs using concentric elliptical bins in the
$(\DeltaM,\ptZ)$ plane.
The bins are chosen in the CRs based on the background distributions and in the SRs based on the signal distributions for each tested signal hypothesis individually, as follows. The angles and proportions of the major and minor axes of the ellipses are obtained by diagonalizing the covariance matrix of \DeltaM and
\ptZ, assuming that \DeltaM and \ptZ are normally distributed. In the CRs, the ellipses are centred around the mean of the total expected background distribution. In the SRs, the ellipses are centred around the mean of the signal distribution and are chosen specifically for each tested signal hypothesis. Six bins are defined in the SRs (CRs) such that the bin boundaries enclose regions of 34, 68, 82, 95, 97, and 99\% of the signal (background) events.
An example of how the elliptical bins are constructed and the respective (\DeltaM, \ptZ) distributions for the signal test hypothesis with $\mA=1000\GeV$ and $\mH=350\GeV$ are shown in Fig.~\ref{fig:ellipses}.
The \finalObs distributions observed in the CRs are shown in
Fig.~\ref{fig:CR:2D}, together with the expected background contributions after
the fit to data described in Section~\ref{sec:results}, for two representative mass hypotheses with large and small \DeltaM.
The data are well described by the background model within the uncertainties.

\section{Systematic uncertainties}\label{sec:systematics}
Several sources of systematic uncertainty alter either the rate or both the rate
and the shape of the \finalObs distributions for the
signal or background processes, and thus
affect the measured signal and background yields.
Effects from the same uncertainty source are treated as fully
correlated across all channels and categories.
Theoretical uncertainties are treated as fully correlated among the
different data-taking periods, while experimental uncertainties are in
most cases treated as uncorrelated between the data recorded in 2016,
2017, and 2018.
The latter assumption is motivated by the fact that the experimental uncertainties are
mainly of statistical origin related to the limited size of the data
and simulation samples used in auxiliary measurements, which are
independent across the data-taking periods.

Since the normalizations of the \ttbar and \ZGjets background
processes are free parameters of the fit to data described in Section~\ref{sec:results}, these are not
affected by the theoretical uncertainties of the inclusive cross section
predictions.
To take into account a possible mismodelling of \ZGHF production, an
additional normalization uncertainty (\ZGHF norm.) of 40\% is assigned to
this process.
The size of the uncertainty was chosen based on results from Ref.~\cite{CMS:2023vzh}, and the results reported in Section~\ref{sec:results} have been confirmed to not depend on the chosen size.
Furthermore, the flexibility of the background model and its robustness against potential mismodelling of the \ZGHF component have been verified using pseudo-experiments.
A large number of pseudo-experiments have been sampled, both from the nominal and from alternative \ZGHF predictions with larger cross sections.
These tests demonstrated that the strength of an injected signal was extracted without bias when fitting with the nominal background model.
Uncertainties in the inclusive cross sections used to
predict the rates of the minor background processes arise primarily
from variations of the renormalization and factorization scales and the PDFs,
and are propagated to the yield estimates.
The cross section uncertainties are each separated into their scale and PDF components, and are correlated where appropriate among processes.
For example, the PDF uncertainties are taken to be correlated among processes with the same predominant initial state, such that they are correlated among \VV and \ttW production.

Uncertainties in the simulation of the signal and background events are evaluated as follows.
Uncertainties arising from missing higher-order terms at the ME level are evaluated by independent variations of the renormalization and factorization scales by factors of 2 and 0.5 with respect to the nominal values.
Two independent nuisance parameters per process are assigned in the fit.
The uncertainty originating from the PDF set is determined from the PDF
variations provided with the {NNPDF3.1} set~\cite{Ball:2017nwa},
correlating processes for which the same flavour scheme and order in the strong coupling constant \alpS are
used in the PDF set.
The uncertainty in the \PYTHIA PS is determined by varying the
parameters controlling the amount of initial- and final-state
radiation independently by factors of 2 and 0.5 with respect to their nominal value~\cite{Sjostrand:2014zea}. Uncertainties due to
the underlying event tune are found to be negligible.
Uncertainties due to the ME-PS matching scheme in the \ttbar
simulation are evaluated by comparing the reference simulation with
samples with varied matching scale parameters.
Their effects on the \DeltaM and \ptZ distributions are found to
affect only the rate of the distributions, and separate rate
variations are assigned per jet multiplicity category, amounting to 6
 and 8\% in the 5j and $\geq$6j categories, respectively.
Uncertainties in the corrections applied to the \pt spectrum of the
\PZ boson in the \ZGjets simulation~\cite{Lindert:2017olm} and of the
\PQt quarks in the \ttbar simulation~\cite{CMS:2018htd} are also
considered.
The former uncertainty ranges from 1 to 10\%, depending on the \ptZ,
while the latter ranges up to 7\% depending on the \pt of the top quarks.

The variations associated with the theory modelling uncertainties described in the previous paragraph
generally affect the overall normalization, the simulated acceptance,
and the shapes of the fitted distributions.
Since the normalization of the signal as well as the \ZGjets and
\ttbar background processes are determined in the fit, the impact
of these variations on the overall normalization before any selection
is removed.
Likewise, for the minor background processes, only the uncertainty in
the inclusive cross section prediction is retained to avoid double
counting of normalization uncertainties.

Effects of the uncertainty in the distribution of the number of pileup
interactions are evaluated by varying the total inelastic cross
section used to predict the number of pileup interactions in the
simulated events by $\pm4.6\%$ from its nominal value~\cite{CMS:2020ebo}.
The impact of statistical fluctuations in the signal and background
prediction due to the limited number of simulated events is accounted
for using the Barlow--Beeston-lite approach~\cite{Conway:2011in}.

The uncertainties in the integrated luminosities are in the 1.2--2.5\% range, depending on the year of data
taking~\cite{CMS-LUM-17-003,CMS-PAS-LUM-17-004,CMS-PAS-LUM-18-002}, and result in an overall uncertainty of 1.6\% for the entire data set,
taking into account the correlations.
The trigger efficiency is measured in an unbiased sample of events, mostly \ttbar leptonic decays, that contain two leptons and fulfill the offline event selection criteria but were collected with triggers based on missing transverse momentum, which feature a negligible correlation with the triggers used in this search.
Corrections are derived to match the efficiency in simulation to that
observed in data, with uncertainties of 1--6\%, dominated by
statistical fluctuations in the data samples used in the measurement.
Additional corrections are applied for a gradual shift in the timing
of the ECAL inputs to the first-level trigger in the region $\abs{\eta} >
2$, which caused a specific trigger inefficiency during the 2016 and
2017 data-taking periods~\cite{CMS:2020cmk}, with uncertainties of approximately
0.5\%.

Residual differences between the lepton reconstruction,
identification, and isolation efficiencies in data and simulation are
corrected based on efficiency measurements using a ``tag-and-probe''
method in data samples of $\PZ\to\Pell\Pell$
events~\cite{CMS:2020uim,CMS:2018rym}.
The uncertainties are smaller than 1 and 2\% for the muon and electron
correction factors, respectively.

The uncertainties in the JES and JER are obtained
by varying the corrections applied to all jets in the
signal and background simulations within their uncertainty.
These uncertainties are divided into nine independent
sources, which include uncertainties due to different
jet flavour compositions in the simulated samples used in the derivation
of the corrections.
While most of the sources are dominated by statistical fluctuations in
auxiliary measurements, some sources are related to theoretical
predictions in the MC simulation and are thus treated as
correlated among the data-taking periods.
The \PQb jet (mis)identification probabilities in simulated events
are corrected to match the efficiencies measured in
data~\cite{BTV-16-002}.
The uncertainties in these corrections, which can arise from
the contamination of background processes in the data samples used for
their derivation, are considered separately for light jets and
\PQb/\PQc jets, and vary between 5 and 30\% depending on the jet type
and \pt~\cite{CMS-DP-2023-005}.

\section{Results}\label{sec:results}
\begin{figure*}[!htp]
  \centering
  \includegraphics[width=0.7\textwidth]{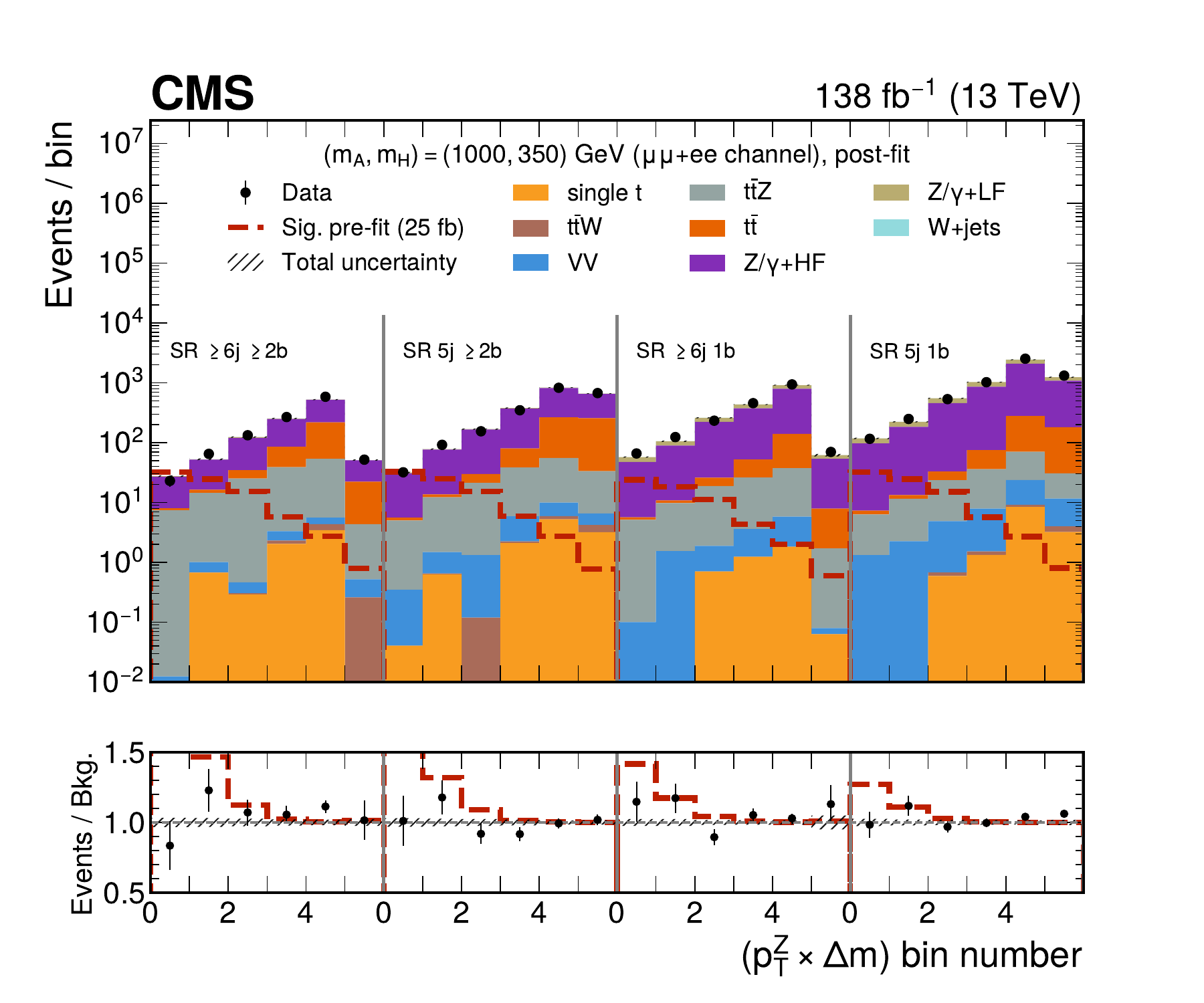}
  \includegraphics[width=0.7\textwidth]{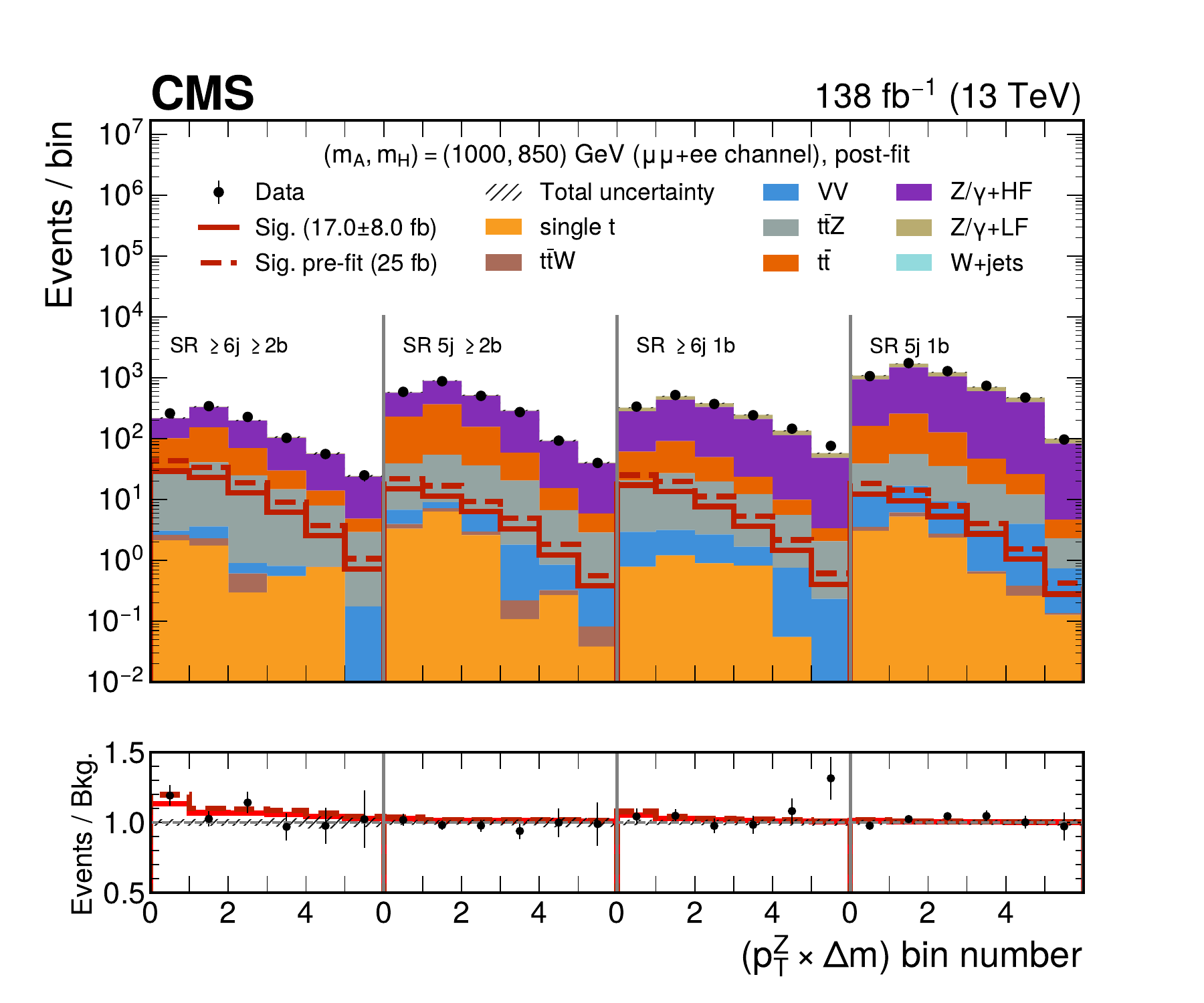}
  \caption{
    Distributions of \finalObs in the SRs after the fit to data with a signal hypothesis of $\mA=1000\GeV$ and $\mH=350$ (upper) and
    $\mH=850\GeV$ (lower).
    The signal (solid red line) and background (coloured histograms)
    distributions are shown with their best fit normalizations from
    the simultaneous fit to the data (``post-fit''), where the sum
    of the \mumu and \ee channels is shown for illustration purposes.
    In the upper plot, the post-fit signal cross section is found to be $0.0\pm0.1\unit{fb}$.
    Therefore, the corresponding histogram is not displayed.
    The signal is also shown for a normalization to 25\unit{fb} (``pre-fit'') in the dashed red line.
    The lower panels show the ratio of the observed data to the background events (black points with error bars) and the ratio of the total number of expected signal-plus-background events to the background events (red lines).
    In both panels, the hatched area represents the total uncertainty.
  }
  \label{fig:results:SR:2D}
\end{figure*}

A simultaneous binned profile likelihood fit of the signal and
background templates to the observed \finalObs distribution in all
channels and categories is performed, using the \textsc{Combine} program~\cite{CMS:2024onh}.
Separate fits are conducted for each $(\mA,\mH)$ hypothesis considered, where the binning of the \finalObs distributions in the SRs has been optimized for each hypothesis, as described in Section~\ref{sec:strategy}.
The mass hypotheses probed range from 430 to 2100\GeV in \mA, and 330 to 2000\GeV in \mH, with steps of 50\GeV at low masses and 100\GeV at higher masses.
The spacing has been chosen based on the expected sensitivity to a potential signal, evaluated using simulation, and taking into account the experimental resolution.
The exact mass values are provided in the HEPData record for this analysis~\cite{HEPData}.
The rates of the \ttbar and \ZGjets backgrounds are free parameters in
the fit, and the systematic uncertainties described in
Section~\ref{sec:systematics} are taken into account via
nuisance parameters in the likelihood function.
Log-normal distributions are used to model the effects of nuisance parameters on the overall rate, while nuisance parameters that affect the shape of the distributions are parametrized using an interpolation between alternative templates~\cite{CMS:2024onh,Conway:2011in}.
Representative \finalObs distributions for $(\mA,\mH)$ hypotheses with
large and small mass splittings in the SRs are shown in Fig.~\ref{fig:results:SR:2D}.

The post-fit values of the background normalization parameters are consistent for all tested mass hypotheses.
The values range between 0.82 and 0.94 with uncertainties of approximately 10\% for \ttbar, and between 0.81 and 0.97 with uncertainties of approximately 14\% for \ZGjets, relative to the nominal predictions.
The observed yield of the \ZGHF background component is approximately 40\% larger than the pre-fit expectation, driven by the combined effect of the \ZGjets normalization parameter together with variations of the nuisance parameters associated with the \ZGHF norm.\ and the renormalization scale uncertainties.
It has been verified that the observed yield of the \ZGHF component does not depend on the size of the \ZGHF norm.\ uncertainty, for example by increasing the \ZGHF norm.\ uncertainty from 40\% to 100\%.
The post-fit values of the other nuisance parameters are
consistent for the different fits and are generally well within 1\SD of
their prior uncertainties.

\begin{figure*}[!htp]
  \centering
  \includegraphics[width=0.49\textwidth]{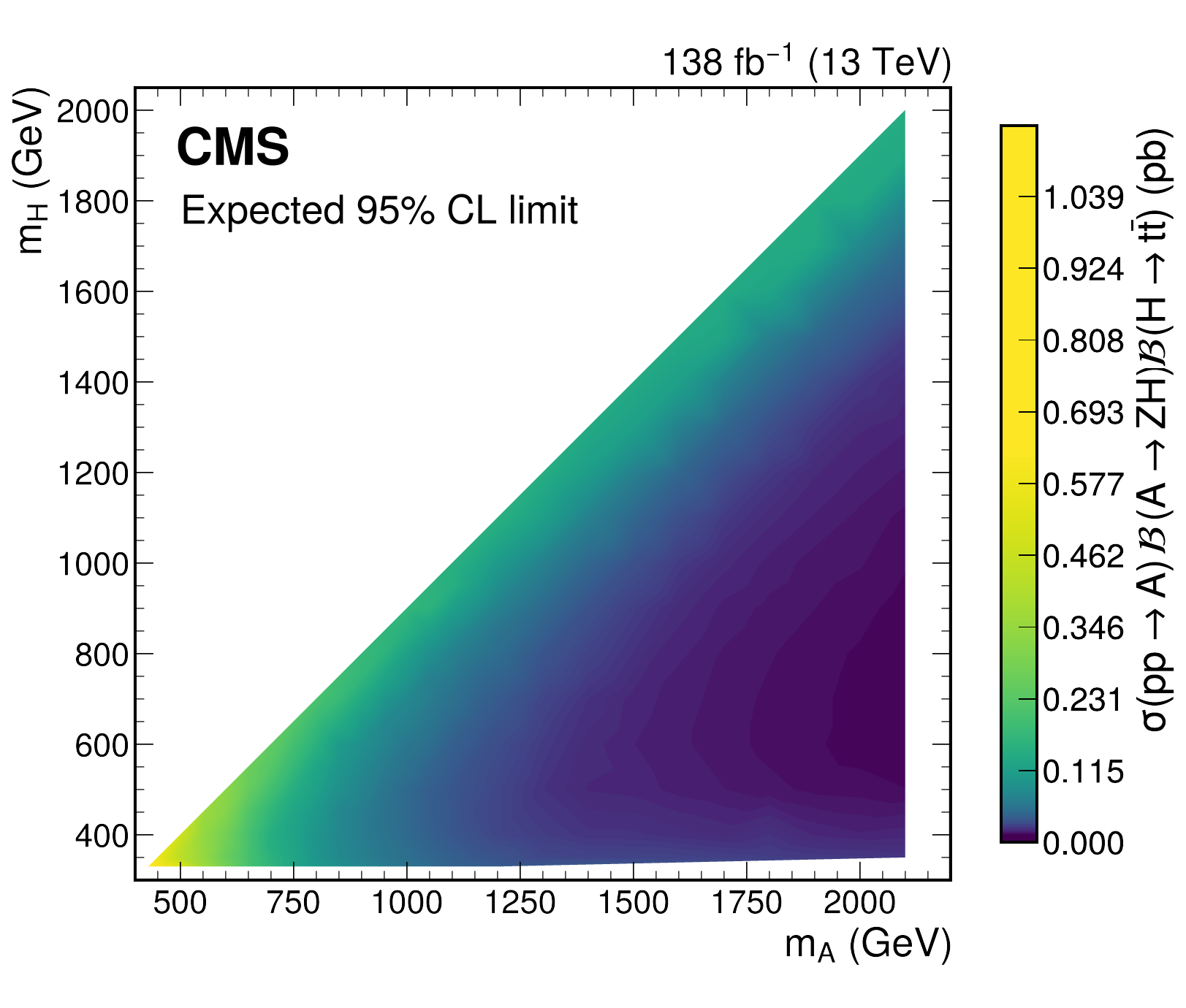}
  \includegraphics[width=0.49\textwidth]{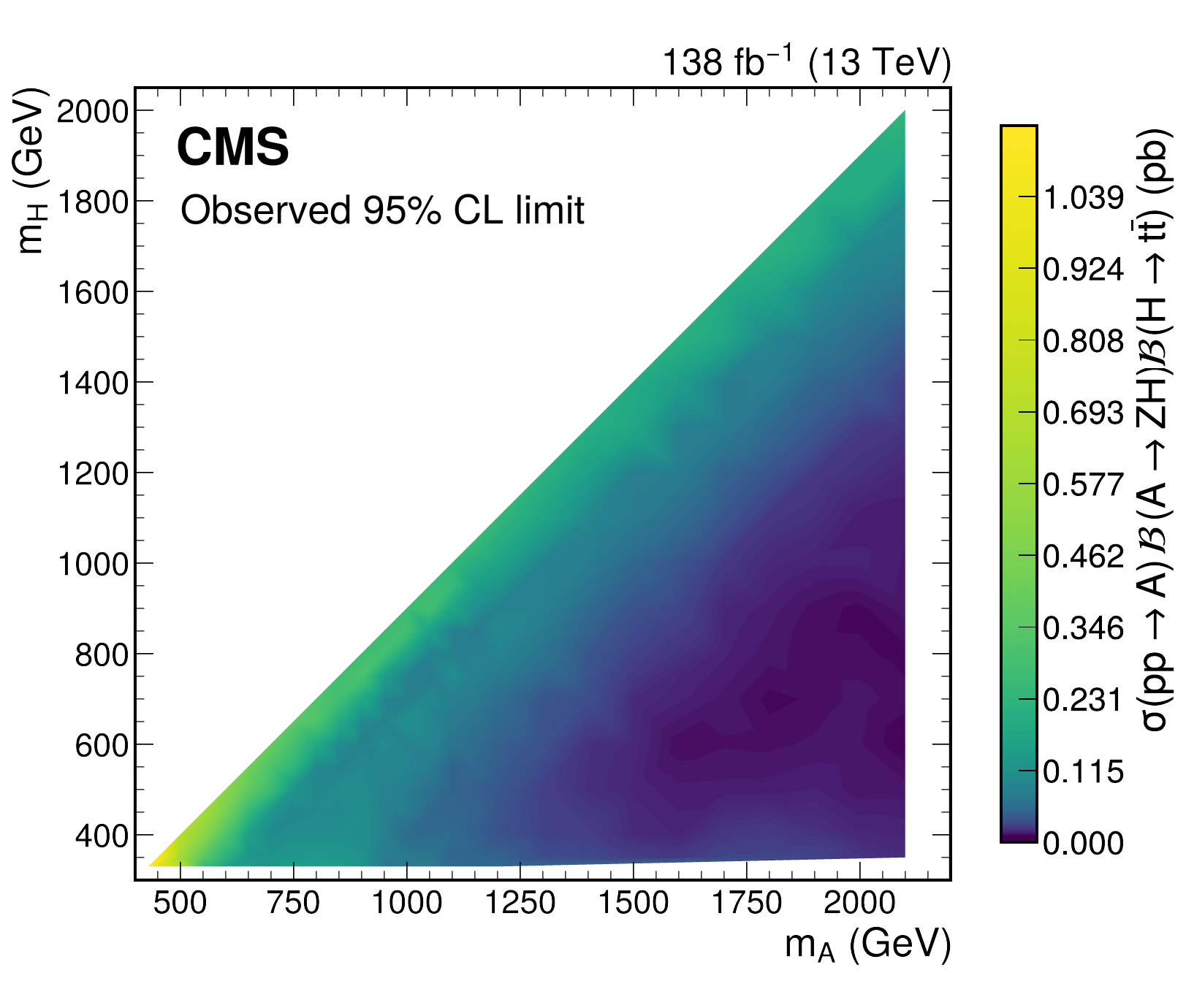}
  \caption{
    Expected (left) and observed (right) 95\% \CL upper limits on the product of production cross section and branching fractions of the \AZHttZ process in the $(\mA, \mH)$ plane.
  }
  \label{fig:xslimits:allpoints:2D}
\end{figure*}

None of the fits show any significant signal excess with respect to the background-only hypothesis.
The largest fluctuation is
observed for $\mA=1000\GeV$ and $\mH=850\GeV$, corresponding to a
local significance of 2.1\SD.
This is visible, for example, in the \mbox{SR $\geq$6j $\geq$2b} category in the lower plot of Fig.~\ref{fig:results:SR:2D}.
Upper limits on the product of cross section and branching fractions, \xsbr, of Higgs-like narrow resonances \PSA and \PSH are derived.
The limits are determined using the \CLs
criterion~\cite{Junk:1999kv,Read:2002hq}, assuming asymptotic distributions~\cite{Cowan:2010js} of the profile likelihood ratio test statistics defined in Ref.~\cite{CMS:2024onh}.
The observed and expected upper limits at 95\% \CL
are presented in Fig.~\ref{fig:xslimits:allpoints:2D} in the $(\mA,
\mH)$ plane and cover the region up to $\mA=2100\GeV$ and $\mH=2000\GeV$.
It has been confirmed using signal injection tests that the search is sensitive to a potential signal at mass configurations between the tested hypotheses, with negligible loss in sensitivity, owing to the finite experimental resolution and the density of the tested points in the $(\mA,\mH)$ plane.
Therefore, the limits are interpolated in Fig.~\ref{fig:xslimits:allpoints:2D} to cover the whole $(\mA,\mH)$ plane.
The excess of events reported by the ATLAS Collaboration in the region around $(\mA,\mH)=(650,450)\GeV$ with a local significance of
2.85\SD~\cite{ATLAS:2023zkt} is not observed in the search presented here, which has a similar expected sensitivity.
The local significance of the deviation from the background at this
point amounts to 0.4\SD, and a \AZHttZ signal with a cross section
larger than 0.25\unit{pb} is excluded at 95\% \CL.

The upper limits on \xsbr are also interpreted in the context
of type-II 2HDMs.
Cross section predictions in the 2HDM are computed with the program \textsc{SusHi}~v.\ 1.7.0~\cite{Harlander:2016hcx} and the branching fractions with \textsc{2HDMC}~v.\ 1.8.0~\cite{Eriksson:2009ws}.
The calculations are performed following Ref.~\cite{Bernon:2015qea}
for different values of \mA, \mH, and \tanb, with $m^{2}_{12}$
set to
\mbox{$\mH^{2}\sin(\beta)\cos(\beta)$} to ensure
vacuum stability over the entire $(\mA,\mH)$ plane.
If not specified otherwise, $\cos(\beta-\alpha)$ is
set to 0, \ie corresponding to the alignment limit.

\begin{figure}[!htp]
  \centering
  \includegraphics[width=0.49\textwidth]{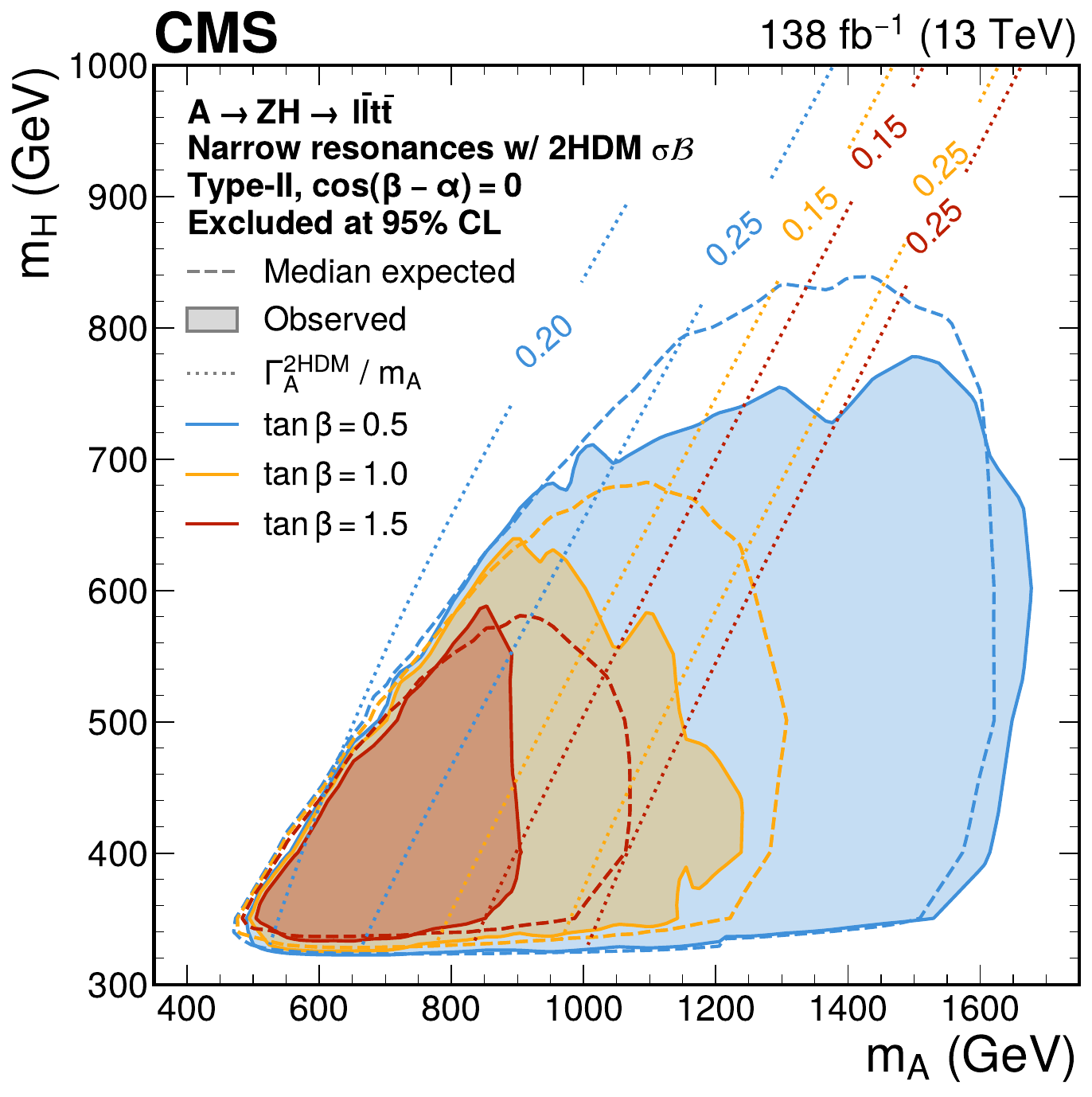}
  \caption{
    Median expected (dashed lines) and observed (filled contours) 95\% \CL
    exclusion regions in the $(\mH,\mA)$ parameter space of the type-II 2HDM for $\tanb = 0.5$ (blue), 1 (orange), and 1.5 (red), derived assuming narrow resonances.
    The enclosed regions are excluded.
    The dotted lines indicate points of constant total decay width \GammaA of the \PSA boson relative to its mass in the 2HDM, for the $\tanb$ value of the corresponding colour.
  }
  \label{fig:2hdm:limits:mHmA}
\end{figure}

\begin{figure*}[!htp]
  \centering
  \includegraphics[width=0.49\textwidth]{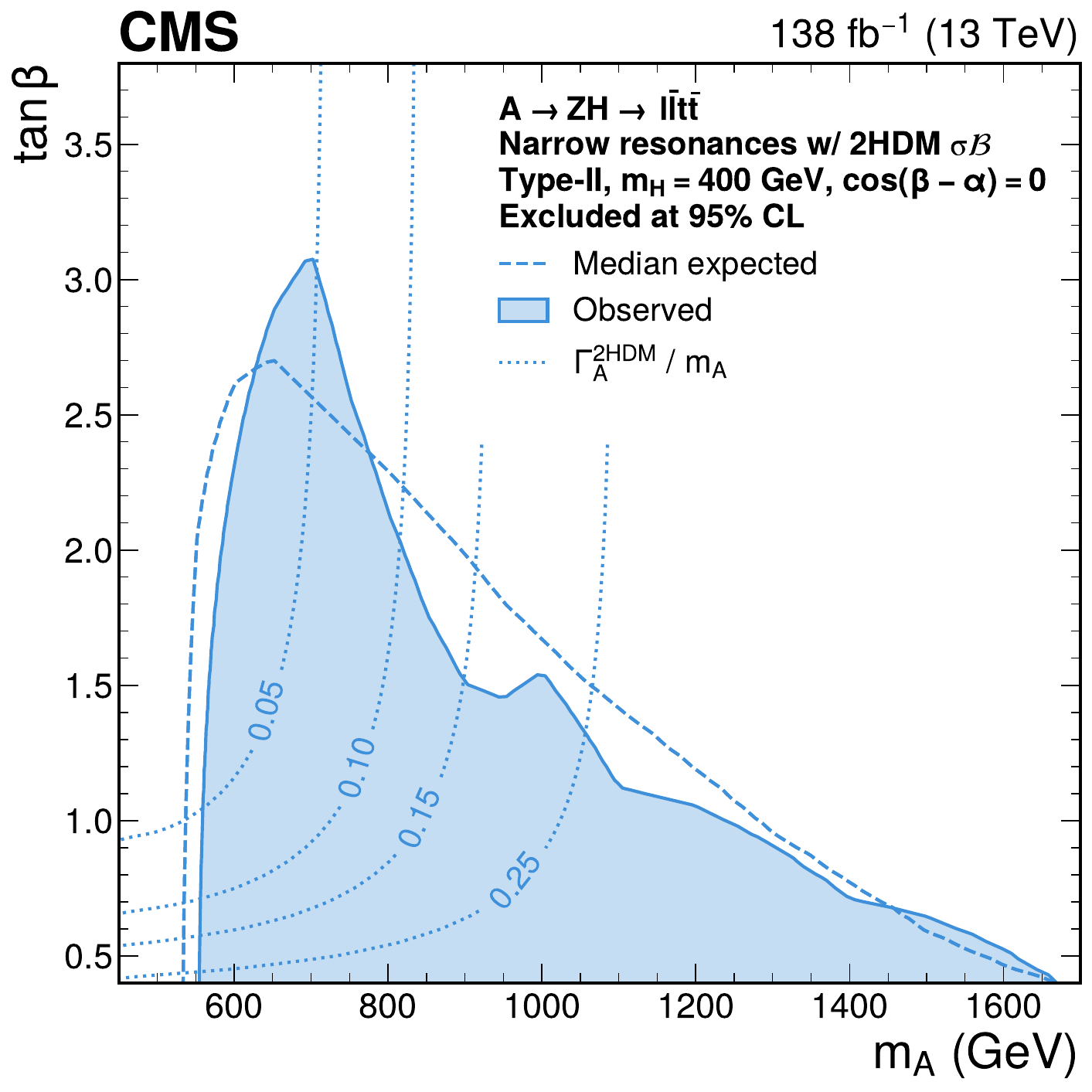}
  \includegraphics[width=0.49\textwidth]{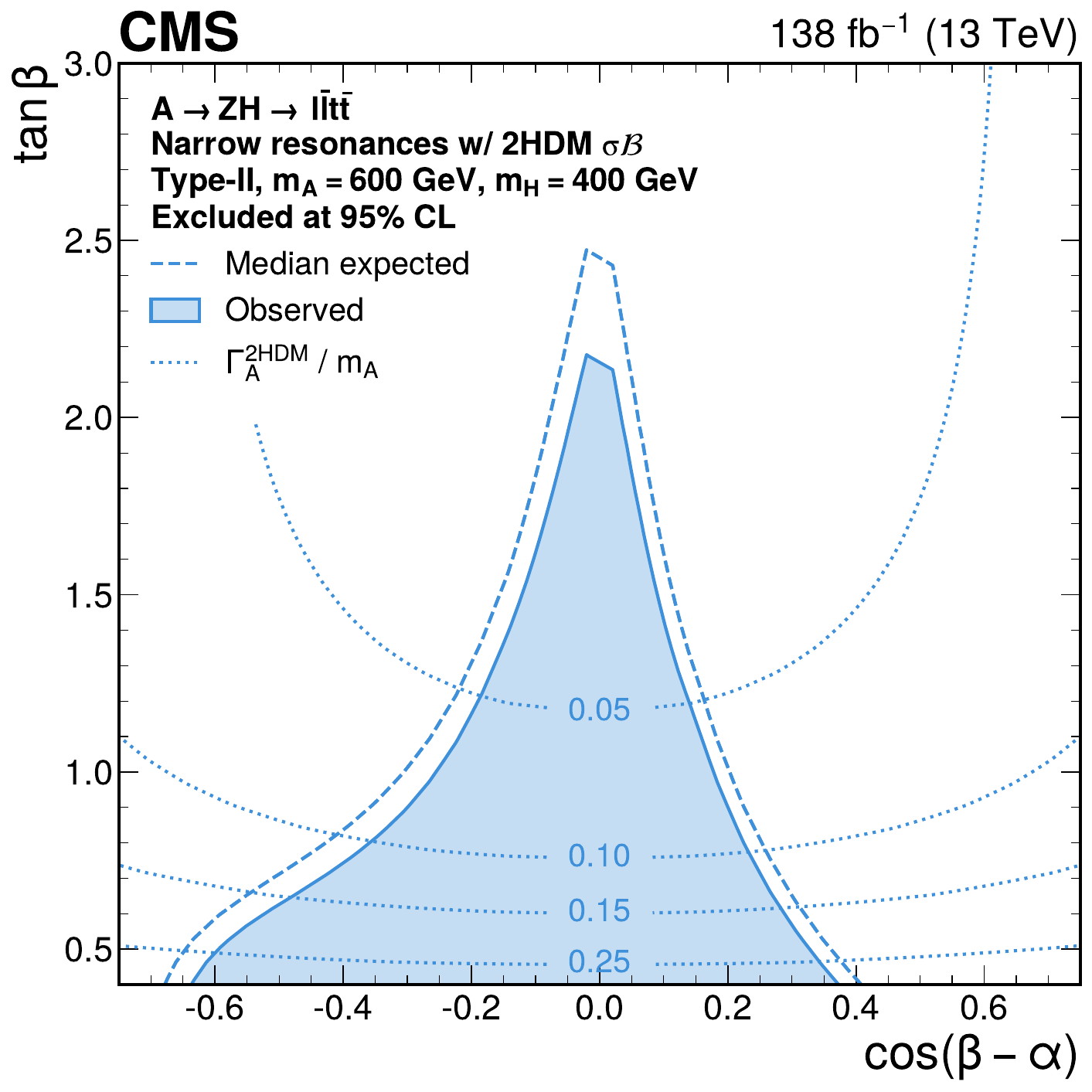}
  \caption{
    Median expected (dashed lines) and observed (filled contours) 95\% \CL
    exclusion regions in the type-II 2HDM $(\tanb, \mA)$ parameter space at $\mH=400\GeV$ (left) and in the $(\tanb,\cos(\beta-\alpha))$ parameter space at $\mA=600\GeV$ and $\mH=400\GeV$ (right), derived assuming narrow resonances.
    The enclosed regions are excluded.
    The dotted lines indicate points of constant total decay width \GammaA of the \PSA boson relative to its mass in the 2HDM.
  }
  \label{fig:2hdm:limits:other}
\end{figure*}

The excluded regions of the 2HDM parameter space are shown in
Fig.~\ref{fig:2hdm:limits:mHmA} as a function of \mA and \mH for
different values of \tanb, assuming narrow resonances.
The results exclude parts of the parameter space relevant for models of electroweak baryogenesis~\cite{Biekotter:2022kgf}.
In Fig.~\ref{fig:2hdm:limits:other}, the excluded regions are shown as a function of \tanb and \mA for $\mH=400\GeV$, and as a function of \tanb and $\cos(\beta-\alpha)$ for $\mA=600\GeV$ and $\mH=400\GeV$, probing the parameter space away from the alignment limit.

The total decay width \GammaA of the \PSA boson in the 2HDM
depends on \mA and \tanb, as shown by the dotted lines in Figs.~\ref{fig:2hdm:limits:mHmA} and~\ref{fig:2hdm:limits:other}.
The width can become sizeable compared to the experimental resolution,
especially for large values of \mA beyond 1\TeV and small values of
\tanb, reducing the sensitivity of the analysis compared to the case
of narrow resonances of the same mass.
The effect has been evaluated using signal events simulated with different values of \GammaA.
For larger signal widths, we verified that the signal significance is reduced by at most 10\% compared to a narrow-resonance signal. This includes all \GammaA values in the 2HDM parameter space considered in this analysis. Thus the narrow-resonance search would be sensitive to a 2HDM signal and the observed absence of a significant excess also constrains this class of models.
For $\tanb=1$, the expected limit on \xsbr is 20\% larger compared to the limit
obtained for narrow resonances at $(\mA,\mH)=(750,330)\GeV$, corresponding to $\GammaA/\mA=15\%$, and 47\% larger at $(1000,400)\GeV$, corresponding to $\GammaA/\mA=25\%$.
The reduction in sensitivity at a given mass point depends on $\tanb$: it is larger for $\tanb=0.5$ than for $\tanb=1$ because $\GammaA$ is larger, and smaller for $\tanb=1.5$ because $\GammaA$ is smaller.

\ifthenelse{\boolean{cms@external}}{}{\clearpage}

\section{Summary}\label{sec:summary}
A direct search for heavy neutral Higgs bosons \PSA and \PSH in the
\ttZ channel has been conducted, using
proton-proton collision data collected by the CMS experiment at $\sqrt{s}=13\TeV$, corresponding to an integrated luminosity of 138\fbinv.
The search targets the process \ppAZH with $\PSH\to\ttbar$.
For the first time, the final state with decays of the \PZ boson to
muons or electrons in association with all-hadronic decays of the \ttbar system
is considered.

No evidence for a signal is observed. Stringent upper limits are set on the product of the cross section and branching fractions for \PSA and \PSH boson masses up to
$\mA=2100\GeV$ and $\mH=2000\GeV$, respectively, assuming narrow resonances.
The results are further used to constrain the parameter space of two-Higgs-doublet models as a function of the \tanb and $\cos(\beta-\alpha)$ parameters.
The excess of events reported by the ATLAS Collaboration in the region around $(\mA,\mH)=(650,450)\GeV$ with a local significance of
2.85 standard deviations~\cite{ATLAS:2023zkt} is not observed in the search presented here, which has similar expected sensitivity.
The results complement and substantially extend the reach of previous
searches, constraining parameter regions relevant for models explaining baryogenesis.

\begin{acknowledgments}
  We congratulate our colleagues in the CERN accelerator departments for the excellent performance of the LHC and thank the technical and administrative staffs at CERN and at other CMS institutes for their contributions to the success of the CMS effort. In addition, we gratefully acknowledge the computing centres and personnel of the Worldwide LHC Computing Grid and other centres for delivering so effectively the computing infrastructure essential to our analyses. Finally, we acknowledge the enduring support for the construction and operation of the LHC, the CMS detector, and the supporting computing infrastructure provided by the following funding agencies: SC (Armenia), BMBWF and FWF (Austria); FNRS and FWO (Belgium); CNPq, CAPES, FAPERJ, FAPERGS, and FAPESP (Brazil); MES and BNSF (Bulgaria); CERN; CAS, MoST, and NSFC (China); MINCIENCIAS (Colombia); MSES and CSF (Croatia); RIF (Cyprus); SENESCYT (Ecuador); ERC PRG, RVTT3 and MoER TK202 (Estonia); Academy of Finland, MEC, and HIP (Finland); CEA and CNRS/IN2P3 (France); SRNSF (Georgia); BMBF, DFG, and HGF (Germany); GSRI (Greece); NKFIH (Hungary); DAE and DST (India); IPM (Iran); SFI (Ireland); INFN (Italy); MSIP and NRF (Republic of Korea); MES (Latvia); LMTLT (Lithuania); MOE and UM (Malaysia); BUAP, CINVESTAV, CONACYT, LNS, SEP, and UASLP-FAI (Mexico); MOS (Montenegro); MBIE (New Zealand); PAEC (Pakistan); MES and NSC (Poland); FCT (Portugal);  MESTD (Serbia); MCIN/AEI and PCTI (Spain); MOSTR (Sri Lanka); Swiss Funding Agencies (Switzerland); MST (Taipei); MHESI and NSTDA (Thailand); TUBITAK and TENMAK (Turkey); NASU (Ukraine); STFC (United Kingdom); DOE and NSF (USA).
   
  \hyphenation{Rachada-pisek} Individuals have received support from the Marie-Curie programme and the European Research Council and Horizon 2020 Grant, contract Nos.\ 675440, 724704, 752730, 758316, 765710, 824093, 101115353, 101002207, and COST Action CA16108 (European Union); the Leventis Foundation; the Alfred P.\ Sloan Foundation; the Alexander von Humboldt Foundation; the Science Committee, project no. 22rl-037 (Armenia); the Belgian Federal Science Policy Office; the Fonds pour la Formation \`a la Recherche dans l'Industrie et dans l'Agriculture (FRIA-Belgium); the F.R.S.-FNRS and FWO (Belgium) under the ``Excellence of Science -- EOS" -- be.h project n.\ 30820817; the Beijing Municipal Science \& Technology Commission, No. Z191100007219010 and Fundamental Research Funds for the Central Universities (China); the Ministry of Education, Youth and Sports (MEYS) of the Czech Republic; the Shota Rustaveli National Science Foundation, grant FR-22-985 (Georgia); the Deutsche Forschungsgemeinschaft (DFG), among others, under Germany's Excellence Strategy -- EXC 2121 ``Quantum Universe" -- 390833306, and under project number 400140256 - GRK2497; the Hellenic Foundation for Research and Innovation (HFRI), Project Number 2288 (Greece); the Hungarian Academy of Sciences, the New National Excellence Program - \'UNKP, the NKFIH research grants K 131991, K 133046, K 138136, K 143460, K 143477, K 146913, K 146914, K 147048, 2020-2.2.1-ED-2021-00181, TKP2021-NKTA-64, and 2021-4.1.2-NEMZ\_KI (Hungary); the Council of Science and Industrial Research, India; ICSC -- National Research Centre for High Performance Computing, Big Data and Quantum Computing and FAIR -- Future Artificial Intelligence Research, funded by the NextGenerationEU program (Italy); the Latvian Council of Science; the Ministry of Education and Science, project no. 2022/WK/14, and the National Science Center, contracts Opus 2021/41/B/ST2/01369 and 2021/43/B/ST2/01552 (Poland); the Funda\c{c}\~ao para a Ci\^encia e a Tecnologia, grant CEECIND/01334/2018 (Portugal); the National Priorities Research Program by Qatar National Research Fund; MCIN/AEI/10.13039/501100011033, ERDF ``a way of making Europe", and the Programa Estatal de Fomento de la Investigaci{\'o}n Cient{\'i}fica y T{\'e}cnica de Excelencia Mar\'{\i}a de Maeztu, grant MDM-2017-0765 and Programa Severo Ochoa del Principado de Asturias (Spain); the Chulalongkorn Academic into Its 2nd Century Project Advancement Project, and the National Science, Research and Innovation Fund via the Program Management Unit for Human Resources \& Institutional Development, Research and Innovation, grant B39G670016 (Thailand); the Kavli Foundation; the Nvidia Corporation; the SuperMicro Corporation; the Welch Foundation, contract C-1845; and the Weston Havens Foundation (USA).
\end{acknowledgments}

\bibliography{auto_generated} 
\cleardoublepage \appendix\section{The CMS Collaboration \label{app:collab}}\begin{sloppypar}\hyphenpenalty=5000\widowpenalty=500\clubpenalty=5000\input{B2G-23-006-public-authorlist.tex}\end{sloppypar}
%%% END EDITABLE REGION %%%
% skeleton_end
\end{document}

%% file: B2G-23-006-public-authorlist.tex
\cmsinstitute{Yerevan Physics Institute, Yerevan, Armenia}
{\tolerance=6000
A.~Hayrapetyan, A.~Tumasyan\cmsAuthorMark{1}\cmsorcid{0009-0000-0684-6742}
\par}
\cmsinstitute{Institut f\"{u}r Hochenergiephysik, Vienna, Austria}
{\tolerance=6000
W.~Adam\cmsorcid{0000-0001-9099-4341}, J.W.~Andrejkovic, L.~Benato\cmsorcid{0000-0001-5135-7489}, T.~Bergauer\cmsorcid{0000-0002-5786-0293}, S.~Chatterjee\cmsorcid{0000-0003-2660-0349}, K.~Damanakis\cmsorcid{0000-0001-5389-2872}, M.~Dragicevic\cmsorcid{0000-0003-1967-6783}, P.S.~Hussain\cmsorcid{0000-0002-4825-5278}, M.~Jeitler\cmsAuthorMark{2}\cmsorcid{0000-0002-5141-9560}, N.~Krammer\cmsorcid{0000-0002-0548-0985}, A.~Li\cmsorcid{0000-0002-4547-116X}, D.~Liko\cmsorcid{0000-0002-3380-473X}, I.~Mikulec\cmsorcid{0000-0003-0385-2746}, J.~Schieck\cmsAuthorMark{2}\cmsorcid{0000-0002-1058-8093}, R.~Sch\"{o}fbeck\cmsAuthorMark{2}\cmsorcid{0000-0002-2332-8784}, D.~Schwarz\cmsorcid{0000-0002-3821-7331}, M.~Sonawane\cmsorcid{0000-0003-0510-7010}, W.~Waltenberger\cmsorcid{0000-0002-6215-7228}, C.-E.~Wulz\cmsAuthorMark{2}\cmsorcid{0000-0001-9226-5812}
% ...
\par}
\vskip\cmsinstskip
\dag:~Deceased\\
$^{1}$Also at Yerevan State University, Yerevan, Armenia\\
$^{2}$Also at TU Wien, Vienna, Austria\\
% ...


%% file: B2G-23-006_temp.bbl
\providecommand{\href}[2]{#2}\begingroup\raggedright\begin{thebibliography}{10}%
\makeatletter
\providecommand{\hrefCMSnoop }[0]{\@secondoftwo}%
\makeatother
\providecommand{\doi}{\texttt{doi:}\begingroup \urlstyle{tt}\Url}

\bibitem{ATLAS:2012yve}
\hrefCMSnoop {}{{ATLAS Collaboration}, ``Observation of a new particle in the
  search for the standard model {Higgs} boson with the {ATLAS} detector at the
  {LHC}'',} \textit{ Phys. Lett. B} \textbf{ 716} (2012) 1,
  \href{http://dx.doi.org/10.1016/j.physletb.2012.08.020}{\doi{10.1016/j.physletb.2012.08.020}},
  \href{http://www.arXiv.org/abs/1207.7214}{\texttt{arXiv:1207.7214}}.

\bibitem{CMS:2012qbp}
\hrefCMSnoop {}{{CMS Collaboration}, ``Observation of a new boson at a mass of
  {$125\GeV$} with the {CMS} experiment at the {LHC}'',} \textit{ Phys. Lett.
  B} \textbf{ 716} (2012) 30,
  \href{http://dx.doi.org/10.1016/j.physletb.2012.08.021}{\doi{10.1016/j.physletb.2012.08.021}},
  \href{http://www.arXiv.org/abs/1207.7235}{\texttt{arXiv:1207.7235}}.

\bibitem{CMS:2013btf}
\hrefCMSnoop {}{{CMS Collaboration}, ``Observation of a new boson with mass
  near {$125\GeV$} in pp collisions at {$\sqrt{s}=7$} and {8\TeV}'',} \textit{
  JHEP} \textbf{ 06} (2013) 081,
  \href{http://dx.doi.org/10.1007/JHEP06(2013)081}{\doi{10.1007/JHEP06(2013)081}},
  \href{http://www.arXiv.org/abs/1303.4571}{\texttt{arXiv:1303.4571}}.

\bibitem{ATLAS:2022vkf}
\hrefCMSnoop {}{{ATLAS Collaboration}, ``A detailed map of {Higgs} boson
  interactions by the {ATLAS} experiment ten years after the discovery'',}
  \textit{ Nature} \textbf{ 607} (2022) 52,
  \href{http://dx.doi.org/10.1038/s41586-022-04893-w}{\doi{10.1038/s41586-022-04893-w}},
  \href{http://www.arXiv.org/abs/2207.00092}{\texttt{arXiv:2207.00092}}.
  [Erratum: \DOI{10.1038/s41586-022-05581-5}].

\bibitem{CMS:2022dwd}
\hrefCMSnoop {}{{CMS Collaboration}, ``A portrait of the {Higgs} boson by the
  {CMS} experiment ten years after the discovery'',} \textit{ Nature} \textbf{
  607} (2022) 60,
  \href{http://dx.doi.org/10.1038/s41586-022-04892-x}{\doi{10.1038/s41586-022-04892-x}},
  \href{http://www.arXiv.org/abs/2207.00043}{\texttt{arXiv:2207.00043}}.

\bibitem{Djouadi:2005gj}
\hrefCMSnoop {}{A.~Djouadi, ``The anatomy of electro-weak symmetry breaking
  {II}: {The} {Higgs} bosons in the minimal supersymmetric model'',} \textit{
  Phys. Rept.} \textbf{ 459} (2008) 1,
  \href{http://dx.doi.org/10.1016/j.physrep.2007.10.005}{\doi{10.1016/j.physrep.2007.10.005}},
  \href{http://www.arXiv.org/abs/hep-ph/0503173}{\texttt{arXiv:hep-ph/0503173}}.

\bibitem{Kim:2008hd}
\hrefCMSnoop {}{J.~E. Kim and G.~Carosi, ``Axions and the strong {CP}
  problem'',} \textit{ Rev. Mod. Phys.} \textbf{ 82} (2010) 557,
  \href{http://dx.doi.org/10.1103/RevModPhys.82.557}{\doi{10.1103/RevModPhys.82.557}},
  \href{http://www.arXiv.org/abs/0807.3125}{\texttt{arXiv:0807.3125}}.
  [Erratum: \DOI{10.1103/RevModPhys.91.049902}].

\bibitem{Bauer:2017ota}
\hrefCMSnoop {}{M.~Bauer, U.~Haisch, and F.~Kahlhoefer, ``Simplified dark
  matter models with two {Higgs} doublets: {I.} pseudoscalar mediators'',}
  \textit{ JHEP} \textbf{ 05} (2017) 138,
  \href{http://dx.doi.org/10.1007/JHEP05(2017)138}{\doi{10.1007/JHEP05(2017)138}},
  \href{http://www.arXiv.org/abs/1701.07427}{\texttt{arXiv:1701.07427}}.

\bibitem{Biekotter:2022kgf}
T.~Biek{\"o}tter\hrefCMSnoop {}{ { et~al.}, ``The trap in the early universe:
  {Impact} on the interplay between gravitational waves and {LHC} physics in
  the {2HDM}'',} \textit{ JCAP} \textbf{ 03} (2023) 031,
  \href{http://dx.doi.org/10.1088/1475-7516/2023/03/031}{\doi{10.1088/1475-7516/2023/03/031}},
  \href{http://www.arXiv.org/abs/2208.14466}{\texttt{arXiv:2208.14466}}.

\bibitem{Lee:1973iz}
\hrefCMSnoop {}{T.~D. Lee, ``A theory of spontaneous {T} violation'',} \textit{
  Phys. Rev. D} \textbf{ 8} (1973) 1226,
  \href{http://dx.doi.org/10.1103/PhysRevD.8.1226}{\doi{10.1103/PhysRevD.8.1226}}.

\bibitem{Branco:2011iw}
G.~C. Branco\hrefCMSnoop {}{ { et~al.}, ``Theory and phenomenology of
  two-{Higgs}-doublet models'',} \textit{ Phys. Rept.} \textbf{ 516} (2012) 1,
  \href{http://dx.doi.org/10.1016/j.physrep.2012.02.002}{\doi{10.1016/j.physrep.2012.02.002}},
  \href{http://www.arXiv.org/abs/1106.0034}{\texttt{arXiv:1106.0034}}.

\bibitem{Haisch:2018djm}
\hrefCMSnoop {}{U.~Haisch and G.~Polesello, ``Searching for heavy {Higgs}
  bosons in the {$\ttbar\PZ$} and {$\ttbar\PW$} final states'',} \textit{ JHEP}
  \textbf{ 09} (2018) 151,
  \href{http://dx.doi.org/10.1007/JHEP09(2018)151}{\doi{10.1007/JHEP09(2018)151}},
  \href{http://www.arXiv.org/abs/1807.07734}{\texttt{arXiv:1807.07734}}.

\bibitem{ATLAS:2020gxx}
\hrefCMSnoop {}{{ATLAS Collaboration}, ``Search for a heavy {Higgs} boson
  decaying into a {\PZ} boson and another heavy {Higgs} boson in the
  {$\Pell\Pell\PQb\PQb$} and {$\Pell\Pell\PW\PW$} final states in {$\Pp\Pp$}
  collisions at $\sqrt{s}=13$\,{TeV} with the {ATLAS} detector'',} \textit{
  Eur. Phys. J. C} \textbf{ 81} (2021) 396,
  \href{http://dx.doi.org/10.1140/epjc/s10052-021-09117-5}{\doi{10.1140/epjc/s10052-021-09117-5}},
  \href{http://www.arXiv.org/abs/2011.05639}{\texttt{arXiv:2011.05639}}.

\bibitem{CMS:2019ogx}
\hrefCMSnoop {}{{CMS Collaboration}, ``Search for new neutral {Higgs} bosons
  through the {$\PH\to\PZ\PSA\to\Pell^{+}\Pell^{-}\bbbar$} process in
  {$\Pp\Pp$} collisions at $\sqrt{s}=13$\,{TeV}'',} \textit{ JHEP} \textbf{ 03}
  (2020) 055,
  \href{http://dx.doi.org/10.1007/JHEP03(2020)055}{\doi{10.1007/JHEP03(2020)055}},
  \href{http://www.arXiv.org/abs/1911.03781}{\texttt{arXiv:1911.03781}}.

\bibitem{CMS:2016xnc}
\hrefCMSnoop {}{{CMS Collaboration}, ``Search for neutral resonances decaying
  into a {\PZ} boson and a pair of {\PQb} jets or {\PGt} leptons'',} \textit{
  Phys. Lett. B} \textbf{ 759} (2016) 369,
  \href{http://dx.doi.org/10.1016/j.physletb.2016.05.087}{\doi{10.1016/j.physletb.2016.05.087}},
  \href{http://www.arXiv.org/abs/1603.02991}{\texttt{arXiv:1603.02991}}.

\bibitem{Dorsch:2013wja}
\hrefCMSnoop {}{G.~C. Dorsch, S.~J. Huber, and J.~M. No, ``A strong electroweak
  phase transition in the {2HDM} after {LHC8}'',} \textit{ JHEP} \textbf{ 10}
  (2013) 029,
  \href{http://dx.doi.org/10.1007/JHEP10(2013)029}{\doi{10.1007/JHEP10(2013)029}},
  \href{http://www.arXiv.org/abs/1305.6610}{\texttt{arXiv:1305.6610}}.

\bibitem{Basler:2016obg}
P.~Basler\hrefCMSnoop {}{ { et~al.}, ``Strong first order electroweak phase
  transition in the {CP}-conserving {2HDM} revisited'',} \textit{ JHEP}
  \textbf{ 02} (2017) 121,
  \href{http://dx.doi.org/10.1007/JHEP02(2017)121}{\doi{10.1007/JHEP02(2017)121}},
  \href{http://www.arXiv.org/abs/1612.04086}{\texttt{arXiv:1612.04086}}.

\bibitem{ATLAS:2024vxm}
\hrefCMSnoop {}{{ATLAS Collaboration}, ``Search for heavy neutral {Higgs}
  bosons decaying into a top quark pair in {$140\fbinv$} of proton-proton
  collision data at $\sqrt{s}=13$\,{TeV} with the {ATLAS} detector'',} \textit{
  JHEP} \textbf{ 08} (2024) 013,
  \href{http://dx.doi.org/10.1007/JHEP08(2024)013}{\doi{10.1007/JHEP08(2024)013}},
  \href{http://www.arXiv.org/abs/2404.18986}{\texttt{arXiv:2404.18986}}.

\bibitem{CMS:2019pzc}
\hrefCMSnoop {}{{CMS Collaboration}, ``Search for heavy {Higgs} bosons decaying
  to a top quark pair in proton-proton collisions at $\sqrt{s}=13$\,{TeV}'',}
  \textit{ JHEP} \textbf{ 04} (2020) 171,
  \href{http://dx.doi.org/10.1007/JHEP04(2020)171}{\doi{10.1007/JHEP04(2020)171}},
  \href{http://www.arXiv.org/abs/1908.01115}{\texttt{arXiv:1908.01115}}.
  [Erratum: \DOI{10.1007/JHEP03(2022)187}].

\bibitem{ATLAS:2018alq}
\hrefCMSnoop {}{{ATLAS Collaboration}, ``Search for new phenomena in events
  with same-charge leptons and {\PQb}-jets in {$\Pp\Pp$} collisions at
  $\sqrt{s}=13$\,{TeV} with the {ATLAS} detector'',} \textit{ JHEP} \textbf{
  12} (2018) 039,
  \href{http://dx.doi.org/10.1007/JHEP12(2018)039}{\doi{10.1007/JHEP12(2018)039}},
  \href{http://www.arXiv.org/abs/1807.11883}{\texttt{arXiv:1807.11883}}.

\bibitem{CMS:2019rvj}
\hrefCMSnoop {}{{CMS Collaboration}, ``Search for production of four top quarks
  in final states with same-sign or multiple leptons in proton-proton
  collisions at $\sqrt{s}=13$\,{TeV}'',} \textit{ Eur. Phys. J. C} \textbf{ 80}
  (2020) 75,
  \href{http://dx.doi.org/10.1140/epjc/s10052-019-7593-7}{\doi{10.1140/epjc/s10052-019-7593-7}},
  \href{http://www.arXiv.org/abs/1908.06463}{\texttt{arXiv:1908.06463}}.

\bibitem{ATLAS:2023zkt}
\hrefCMSnoop {}{{ATLAS Collaboration}, ``Search for a {CP}-odd {Higgs} boson
  decaying to a heavy {CP}-even {Higgs} boson and a {\PZ} boson in the
  {$\Pell\Pell\ttbar$} and {$\PGn\PAGn\bbbar$} final states using
  {$140\,\text{fb}^{-1}$} of data collected with the {ATLAS} detector'',}
  \textit{ JHEP} \textbf{ 02} (2024) 197,
  \href{http://dx.doi.org/10.1007/JHEP02(2024)197}{\doi{10.1007/JHEP02(2024)197}},
  \href{http://www.arXiv.org/abs/2311.04033}{\texttt{arXiv:2311.04033}}.

\bibitem{CMS-LUM-17-003}
\hrefCMSnoop {}{{CMS Collaboration}, ``Precision luminosity measurement in
  proton-proton collisions at $\sqrt{s}=13$\,{TeV} in 2015 and 2016 at
  {CMS}'',} \textit{ Eur. Phys. J. C} \textbf{ 81} (2021) 800,
  \href{http://dx.doi.org/10.1140/epjc/s10052-021-09538-2}{\doi{10.1140/epjc/s10052-021-09538-2}},
  \href{http://www.arXiv.org/abs/2104.01927}{\texttt{arXiv:2104.01927}}.

\bibitem{CMS-PAS-LUM-17-004}
\href {{http://cds.cern.ch/record/2621960}}{{CMS Collaboration}, ``{CMS}
  luminosity measurement for the 2017 data-taking period at
  $\sqrt{s}=13$\,{TeV}'',} {CMS Physics Analysis Summary} CMS-PAS-LUM-17-004,
  2017.

\bibitem{CMS-PAS-LUM-18-002}
\href {{http://cds.cern.ch/record/2676164}}{{CMS Collaboration}, ``{CMS}
  luminosity measurement for the 2018 data-taking period at
  $\sqrt{s}=13$\,{TeV}'',} {CMS Physics Analysis Summary} CMS-PAS-LUM-18-002,
  2018.

\bibitem{HEPData}
\hrefCMSnoop {}{}{HEPD}ata record for this analysis, 2024.
\newblock
  \href{http://dx.doi.org/10.17182/hepdata.150023}{\doi{10.17182/hepdata.150023}}.

\bibitem{CMS:2008xjf}
\hrefCMSnoop {}{{CMS Collaboration}, ``The {CMS} experiment at the {CERN}
  {LHC}'',} \textit{ JINST} \textbf{ 3} (2008) S08004,
  \href{http://dx.doi.org/10.1088/1748-0221/3/08/S08004}{\doi{10.1088/1748-0221/3/08/S08004}}.

\bibitem{CMS:2023gfb}
\hrefCMSnoop {}{{CMS Collaboration}, ``Development of the {CMS} detector for
  the {CERN LHC Run 3}'',} \textit{ JINST} \textbf{ 19} (2024) P05064,
  \href{http://dx.doi.org/10.1088/1748-0221/19/05/P05064}{\doi{10.1088/1748-0221/19/05/P05064}}.

\bibitem{CMS:2020cmk}
\hrefCMSnoop {}{{CMS Collaboration}, ``Performance of the {CMS} level-1 trigger
  in proton-proton collisions at $\sqrt{s}=13$\,{TeV}'',} \textit{ JINST}
  \textbf{ 15} (2020) P10017,
  \href{http://dx.doi.org/10.1088/1748-0221/15/10/P10017}{\doi{10.1088/1748-0221/15/10/P10017}},
  \href{http://www.arXiv.org/abs/2006.10165}{\texttt{arXiv:2006.10165}}.

\bibitem{CMS:2016ngn}
\hrefCMSnoop {}{{CMS Collaboration}, ``The {CMS} trigger system'',} \textit{
  JINST} \textbf{ 12} (2017) P01020,
  \href{http://dx.doi.org/10.1088/1748-0221/12/01/P01020}{\doi{10.1088/1748-0221/12/01/P01020}},
  \href{http://www.arXiv.org/abs/1609.02366}{\texttt{arXiv:1609.02366}}.

\bibitem{CMS:2024aqx}
\hrefCMSnoop {}{{CMS Collaboration}, ``Performance of the {CMS} high-level
  trigger during {LHC Run 2}'',} \textit{ JINST} \textbf{ 19} (2024) P11021,
  \href{http://dx.doi.org/10.1088/1748-0221/19/11/P11021}{\doi{10.1088/1748-0221/19/11/P11021}},
  \href{http://www.arXiv.org/abs/2410.17038}{\texttt{arXiv:2410.17038}}.

\bibitem{CMS:2020uim}
\hrefCMSnoop {}{{CMS Collaboration}, ``Electron and photon reconstruction and
  identification with the {CMS} experiment at the {CERN} {LHC}'',} \textit{
  JINST} \textbf{ 16} (2021) P05014,
  \href{http://dx.doi.org/10.1088/1748-0221/16/05/P05014}{\doi{10.1088/1748-0221/16/05/P05014}},
  \href{http://www.arXiv.org/abs/2012.06888}{\texttt{arXiv:2012.06888}}.

\bibitem{CMS:2018rym}
\hrefCMSnoop {}{{CMS Collaboration}, ``Performance of the {CMS} muon detector
  and muon reconstruction with proton-proton collisions at
  $\sqrt{s}=13$\,{TeV}'',} \textit{ JINST} \textbf{ 13} (2018) P06015,
  \href{http://dx.doi.org/10.1088/1748-0221/13/06/P06015}{\doi{10.1088/1748-0221/13/06/P06015}},
  \href{http://www.arXiv.org/abs/1804.04528}{\texttt{arXiv:1804.04528}}.

\bibitem{CMS:2014pgm}
\hrefCMSnoop {}{{CMS Collaboration}, ``Description and performance of track and
  primary-vertex reconstruction with the {CMS} tracker'',} \textit{ JINST}
  \textbf{ 9} (2014) P10009,
  \href{http://dx.doi.org/10.1088/1748-0221/9/10/P10009}{\doi{10.1088/1748-0221/9/10/P10009}},
  \href{http://www.arXiv.org/abs/1405.6569}{\texttt{arXiv:1405.6569}}.

\bibitem{CMS:2017yfk}
\hrefCMSnoop {}{{CMS Collaboration}, ``Particle-flow reconstruction and global
  event description with the {CMS} detector'',} \textit{ JINST} \textbf{ 12}
  (2017) P10003,
  \href{http://dx.doi.org/10.1088/1748-0221/12/10/P10003}{\doi{10.1088/1748-0221/12/10/P10003}},
\href{http://www.arXiv.org/abs/1706.04965}{\texttt{arXiv:1706.04965}}.
%%CITATION = ARXIV:1706.04965;%%.

\bibitem{CMS:2018jrd}
\hrefCMSnoop {}{{CMS Collaboration}, ``Performance of reconstruction and
  identification of {\PGt} leptons decaying to hadrons and {\PGnGt} in
  {$\Pp\Pp$} collisions at {$\sqrt{s}=13\TeV$}'',} \textit{ JINST} \textbf{ 13}
  (2018) P10005,
  \href{http://dx.doi.org/10.1088/1748-0221/13/10/P10005}{\doi{10.1088/1748-0221/13/10/P10005}},
  \href{http://www.arXiv.org/abs/1809.02816}{\texttt{arXiv:1809.02816}}.

\bibitem{CMS:2016lmd}
\hrefCMSnoop {}{{CMS Collaboration}, ``Jet energy scale and resolution in the
  {CMS} experiment in {$\Pp\Pp$} collisions at 8\,{TeV}'',} \textit{ JINST}
  \textbf{ 12} (2017) P02014,
  \href{http://dx.doi.org/10.1088/1748-0221/12/02/P02014}{\doi{10.1088/1748-0221/12/02/P02014}},
  \href{http://www.arXiv.org/abs/1607.03663}{\texttt{arXiv:1607.03663}}.

\bibitem{CMS:2019ctu}
\hrefCMSnoop {}{{CMS Collaboration}, ``Performance of missing transverse
  momentum reconstruction in proton-proton collisions at {$\sqrt{s}= 13\TeV$}
  using the {CMS} detector'',} \textit{ JINST} \textbf{ 14} (2019) P07004,
  \href{http://dx.doi.org/10.1088/1748-0221/14/07/P07004}{\doi{10.1088/1748-0221/14/07/P07004}},
\href{http://www.arXiv.org/abs/1903.06078}{\texttt{arXiv:1903.06078}}.
%%CITATION = ARXIV:1903.06078;%%.

\bibitem{CMS-TDR-15-02}
\href {http://cds.cern.ch/record/2020886}{{CMS Collaboration}, ``Technical
  proposal for the {Phase-2} upgrade of the {Compact Muon Solenoid}'',} {CMS
  Technical Proposal} CERN-LHCC-2015-010, CMS-TDR-15-02, 2015.

\bibitem{CMS-DP-2020-021}
\href {https://cds.cern.ch/record/2717925}{{CMS Collaboration}, ``{ECAL} 2016
  refined calibration and {Run 2} summary plots'',} {CMS Detector Performance
  Summary} CMS-DP-2020-021, 2020.

\bibitem{CMS:2020ebo}
\hrefCMSnoop {}{{CMS Collaboration}, ``Pileup mitigation at {CMS} in {13\,TeV}
  data'',} \textit{ JINST} \textbf{ 15} (2020) P09018,
  \href{http://dx.doi.org/10.1088/1748-0221/15/09/P09018}{\doi{10.1088/1748-0221/15/09/P09018}},
  \href{http://www.arXiv.org/abs/2003.00503}{\texttt{arXiv:2003.00503}}.

\bibitem{Cacciari:2008gp}
\hrefCMSnoop {}{M.~Cacciari, G.~P. Salam, and G.~Soyez, ``The anti-{$\kt$} jet
  clustering algorithm'',} \textit{ JHEP} \textbf{ 04} (2008) 063,
  \href{http://dx.doi.org/10.1088/1126-6708/2008/04/063}{\doi{10.1088/1126-6708/2008/04/063}},
  \href{http://www.arXiv.org/abs/0802.1189}{\texttt{arXiv:0802.1189}}.

\bibitem{Cacciari:2011ma}
\hrefCMSnoop {}{M.~Cacciari, G.~P. Salam, and G.~Soyez, ``{FastJet} user
  manual'',} \textit{ Eur. Phys. J. C} \textbf{ 72} (2012) 1896,
  \href{http://dx.doi.org/10.1140/epjc/s10052-012-1896-2}{\doi{10.1140/epjc/s10052-012-1896-2}},
  \href{http://www.arXiv.org/abs/1111.6097}{\texttt{arXiv:1111.6097}}.

\bibitem{Cacciari_2008}
\hrefCMSnoop {}{M.~Cacciari, G.~P. Salam, and G.~Soyez, ``The catchment area of
  jets'',} \textit{ JHEP} \textbf{ 04} (2008) 005,
  \href{http://dx.doi.org/10.1088/1126-6708/2008/04/005}{\doi{10.1088/1126-6708/2008/04/005}},
  \href{http://www.arXiv.org/abs/0802.1188}{\texttt{arXiv:0802.1188}}.

\bibitem{BTV-16-002}
\hrefCMSnoop {}{{CMS Collaboration}, ``Identification of heavy-flavour jets
  with the {CMS} detector in {$\Pp\Pp$} collisions at 13\,{TeV}'',} \textit{
  JINST} \textbf{ 13} (2018) P05011,
  \href{http://dx.doi.org/10.1088/1748-0221/13/05/P05011}{\doi{10.1088/1748-0221/13/05/P05011}},
  \href{http://www.arXiv.org/abs/1712.07158}{\texttt{arXiv:1712.07158}}.

\bibitem{Bols:2020bkb}
E.~Bols\hrefCMSnoop {}{ { et~al.}, ``Jet flavour classification using
  {DeepJet}'',} \textit{ JINST} \textbf{ 15} (2020) P12012,
  \href{http://dx.doi.org/10.1088/1748-0221/15/12/P12012}{\doi{10.1088/1748-0221/15/12/P12012}},
  \href{http://www.arXiv.org/abs/2008.10519}{\texttt{arXiv:2008.10519}}.

\bibitem{CMS-DP-2023-005}
\href {https://cds.cern.ch/record/2854609}{{CMS Collaboration}, ``Performance
  summary of {AK4} jet {\PQb} tagging with data from proton-proton collisions
  at 13\,{TeV} with the {CMS} detector'',} {CMS Detector Performance Summary}
  CMS-DP-2023-005, 2023.

\bibitem{Agostinelli:2002hh}
\hrefCMSnoop {}{{GEANT4} Collaboration, ``{\GEANTfour}---a simulation
  toolkit'',} \textit{ Nucl. Instrum. Meth. A} \textbf{ 506} (2003) 250,
  \href{http://dx.doi.org/10.1016/S0168-9002(03)01368-8}{\doi{10.1016/S0168-9002(03)01368-8}}.

\bibitem{Ball:2017nwa}
\hrefCMSnoop {}{{NNPDF} Collaboration, ``Parton distributions from
  high-precision collider data'',} \textit{ Eur. Phys. J. C} \textbf{ 77}
  (2017) 663,
  \href{http://dx.doi.org/10.1140/epjc/s10052-017-5199-5}{\doi{10.1140/epjc/s10052-017-5199-5}},
  \href{http://www.arXiv.org/abs/1706.00428}{\texttt{arXiv:1706.00428}}.

\bibitem{Sjostrand:2014zea}
T.~Sj{\"o}strand\hrefCMSnoop {}{ { et~al.}, ``An introduction to {\PYTHIA}
  $8.2$'',} \textit{ Comput. Phys. Commun.} \textbf{ 191} (2015) 159,
  \href{http://dx.doi.org/10.1016/j.cpc.2015.01.024}{\doi{10.1016/j.cpc.2015.01.024}},
  \href{http://www.arXiv.org/abs/1410.3012}{\texttt{arXiv:1410.3012}}.

\bibitem{Sirunyan:2019dfx}
\hrefCMSnoop {}{{CMS Collaboration}, ``Extraction and validation of a new set
  of {CMS} {\PYTHIA} 8 tunes from underlying-event measurements'',} \textit{
  Eur. Phys. J. C} \textbf{ 80} (2020) 4,
  \href{http://dx.doi.org/10.1140/epjc/s10052-019-7499-4}{\doi{10.1140/epjc/s10052-019-7499-4}},
  \href{http://www.arXiv.org/abs/1903.12179}{\texttt{arXiv:1903.12179}}.

\bibitem{Alwall:2014hca}
J.~Alwall\hrefCMSnoop {}{ { et~al.}, ``The automated computation of tree-level
  and next-to-leading order differential cross sections, and their matching to
  parton shower simulations'',} \textit{ JHEP} \textbf{ 07} (2014) 079,
  \href{http://dx.doi.org/10.1007/JHEP07(2014)079}{\doi{10.1007/JHEP07(2014)079}},
  \href{http://www.arXiv.org/abs/1405.0301}{\texttt{arXiv:1405.0301}}.

\bibitem{Degrande:2014vpa}
\hrefCMSnoop {}{C.~Degrande, ``Automatic evaluation of {UV} and {R2} terms for
  beyond the standard model {Lagrangians}: {A} proof-of-principle'',} \textit{
  Comput. Phys. Commun.} \textbf{ 197} (2015) 239,
  \href{http://dx.doi.org/10.1016/j.cpc.2015.08.015}{\doi{10.1016/j.cpc.2015.08.015}},
  \href{http://www.arXiv.org/abs/1406.3030}{\texttt{arXiv:1406.3030}}.

\bibitem{Degrande:2011ua}
C.~Degrande\hrefCMSnoop {}{ { et~al.}, ``{UFO} -- the {Universal} {FeynRules}
  {Output}'',} \textit{ Comput. Phys. Commun.} \textbf{ 183} (2012) 1201,
  \href{http://dx.doi.org/10.1016/j.cpc.2012.01.022}{\doi{10.1016/j.cpc.2012.01.022}},
  \href{http://www.arXiv.org/abs/1108.2040}{\texttt{arXiv:1108.2040}}.

\bibitem{Nason:2004rx}
\hrefCMSnoop {}{P.~Nason, ``A new method for combining {NLO} {QCD} with shower
  {Monte} {Carlo} algorithms'',} \textit{ JHEP} \textbf{ 11} (2004) 040,
  \href{http://dx.doi.org/10.1088/1126-6708/2004/11/040}{\doi{10.1088/1126-6708/2004/11/040}},
  \href{http://www.arXiv.org/abs/hep-ph/0409146}{\texttt{arXiv:hep-ph/0409146}}.

\bibitem{Frixione:2007vw}
\hrefCMSnoop {}{S.~Frixione, P.~Nason, and C.~Oleari, ``Matching {NLO} {QCD}
  computations with parton shower simulations: {The} {POWHEG} method'',}
  \textit{ JHEP} \textbf{ 11} (2007) 070,
  \href{http://dx.doi.org/10.1088/1126-6708/2007/11/070}{\doi{10.1088/1126-6708/2007/11/070}},
  \href{http://www.arXiv.org/abs/0709.2092}{\texttt{arXiv:0709.2092}}.

\bibitem{Alioli:2010xd}
\hrefCMSnoop {}{S.~Alioli, P.~Nason, C.~Oleari, and E.~Re, ``A general
  framework for implementing {NLO} calculations in shower {Monte} {Carlo}
  programs: {The} {POWHEG BOX}'',} \textit{ JHEP} \textbf{ 06} (2010) 043,
  \href{http://dx.doi.org/10.1007/JHEP06(2010)043}{\doi{10.1007/JHEP06(2010)043}},
  \href{http://www.arXiv.org/abs/1002.2581}{\texttt{arXiv:1002.2581}}.

\bibitem{Jezo:2015aia}
\hrefCMSnoop {}{T.~Je{\v{z}}o and P.~Nason, ``On the treatment of resonances in
  next-to-leading order calculations matched to a parton shower'',} \textit{
  JHEP} \textbf{ 12} (2015) 065,
  \href{http://dx.doi.org/10.1007/JHEP12(2015)065}{\doi{10.1007/JHEP12(2015)065}},
  \href{http://www.arXiv.org/abs/1509.09071}{\texttt{arXiv:1509.09071}}.

\bibitem{Frixione:2007nw}
\hrefCMSnoop {}{S.~Frixione, P.~Nason, and G.~Ridolfi, ``A positive-weight
  next-to-leading-order {Monte} {Carlo} for heavy flavour hadroproduction'',}
  \textit{ JHEP} \textbf{ 09} (2007) 126,
  \href{http://dx.doi.org/10.1088/1126-6708/2007/09/126}{\doi{10.1088/1126-6708/2007/09/126}},
  \href{http://www.arXiv.org/abs/0707.3088}{\texttt{arXiv:0707.3088}}.

\bibitem{bib:ttbarxs:1}
\hrefCMSnoop {}{M.~Beneke, P.~Falgari, S.~Klein, and C.~Schwinn, ``Hadronic
  top-quark pair production with {NNLL} threshold resummation'',} \textit{
  Nucl. Phys. B} \textbf{ 855} (2012) 695,
  \href{http://dx.doi.org/10.1016/j.nuclphysb.2011.10.021}{\doi{10.1016/j.nuclphysb.2011.10.021}},
  \href{http://www.arXiv.org/abs/1109.1536}{\texttt{arXiv:1109.1536}}.

\bibitem{bib:ttbarxs:2}
M.~Cacciari\hrefCMSnoop {}{ { et~al.}, ``Top-pair production at hadron
  colliders with next-to-next-to-leading logarithmic soft-gluon resummation'',}
  \textit{ Phys. Lett. B} \textbf{ 710} (2012) 612,
  \href{http://dx.doi.org/10.1016/j.physletb.2012.03.013}{\doi{10.1016/j.physletb.2012.03.013}},
  \href{http://www.arXiv.org/abs/1111.5869}{\texttt{arXiv:1111.5869}}.

\bibitem{bib:ttbarxs:3}
\hrefCMSnoop {}{P.~B{\"a}rnreuther, M.~Czakon, and A.~Mitov,
  ``Percent-level-precision physics at the {Tevatron}:
  {Next-to-next-to-leading} order {QCD} corrections to
  {$\PQq\PAQq\to\ttbar\text{+X}$}'',} \textit{ Phys. Rev. Lett.} \textbf{ 109}
  (2012) 132001,
  \href{http://dx.doi.org/10.1103/PhysRevLett.109.132001}{\doi{10.1103/PhysRevLett.109.132001}},
  \href{http://www.arXiv.org/abs/1204.5201}{\texttt{arXiv:1204.5201}}.

\bibitem{bib:ttbarxs:4}
\hrefCMSnoop {}{M.~Czakon and A.~Mitov, ``{NNLO} corrections to top-pair
  production at hadron colliders: {The} all-fermionic scattering channels'',}
  \textit{ JHEP} \textbf{ 12} (2012) 054,
  \href{http://dx.doi.org/10.1007/JHEP12(2012)054}{\doi{10.1007/JHEP12(2012)054}},
  \href{http://www.arXiv.org/abs/1207.0236}{\texttt{arXiv:1207.0236}}.

\bibitem{bib:ttbarxs:5}
\hrefCMSnoop {}{M.~Czakon and A.~Mitov, ``{NNLO} corrections to top pair
  production at hadron colliders: {The} quark-gluon reaction'',} \textit{ JHEP}
  \textbf{ 01} (2013) 080,
  \href{http://dx.doi.org/10.1007/JHEP01(2013)080}{\doi{10.1007/JHEP01(2013)080}},
  \href{http://www.arXiv.org/abs/1210.6832}{\texttt{arXiv:1210.6832}}.

\bibitem{bib:ttbarxs:6}
\hrefCMSnoop {}{M.~Czakon, P.~Fiedler, and A.~Mitov, ``Total top-quark
  pair-production cross section at hadron colliders through
  {$\mathcal{O}(\alpha_s^4)$}'',} \textit{ Phys. Rev. Lett.} \textbf{ 110}
  (2013) 252004,
  \href{http://dx.doi.org/10.1103/PhysRevLett.110.252004}{\doi{10.1103/PhysRevLett.110.252004}},
  \href{http://www.arXiv.org/abs/1303.6254}{\texttt{arXiv:1303.6254}}.

\bibitem{bib:ttbarxs:7}
\hrefCMSnoop {}{M.~Czakon and A.~Mitov, ``{Top++}: {A} program for the
  calculation of the top-pair cross-section at hadron colliders'',} \textit{
  Comput. Phys. Commun.} \textbf{ 185} (2014) 2930,
  \href{http://dx.doi.org/10.1016/j.cpc.2014.06.021}{\doi{10.1016/j.cpc.2014.06.021}},
  \href{http://www.arXiv.org/abs/1112.5675}{\texttt{arXiv:1112.5675}}.

\bibitem{CMS:2018htd}
\hrefCMSnoop {}{{CMS Collaboration}, ``Measurement of differential cross
  sections for the production of top quark pairs and of additional jets in
  lepton+jets events from {$\Pp\Pp$} collisions at $\sqrt{s}=13$\,{TeV}'',}
  \textit{ Phys. Rev. D} \textbf{ 97} (2018) 112003,
  \href{http://dx.doi.org/10.1103/PhysRevD.97.112003}{\doi{10.1103/PhysRevD.97.112003}},
  \href{http://www.arXiv.org/abs/1803.08856}{\texttt{arXiv:1803.08856}}.

\bibitem{CMS:2024ybg}
\hrefCMSnoop {}{{CMS Collaboration}, ``Differential cross section measurements
  for the production of top quark pairs and of additional jets using dilepton
  events from {$\Pp\Pp$} collisions at $\sqrt{s}=13$\,{TeV}'',} \textit{ JHEP}
  \textbf{ 02} (2025) 064,
  \href{http://dx.doi.org/10.1007/JHEP02(2025)064}{\doi{10.1007/JHEP02(2025)064}},
  \href{http://www.arXiv.org/abs/2402.08486}{\texttt{arXiv:2402.08486}}.

\bibitem{Alwall:2007fs}
J.~Alwall\hrefCMSnoop {}{ { et~al.}, ``Comparative study of various algorithms
  for the merging of parton showers and matrix elements in hadronic
  collisions'',} \textit{ Eur. Phys. J. C} \textbf{ 53} (2008) 473,
  \href{http://dx.doi.org/10.1140/epjc/s10052-007-0490-5}{\doi{10.1140/epjc/s10052-007-0490-5}},
  \href{http://www.arXiv.org/abs/0706.2569}{\texttt{arXiv:0706.2569}}.

\bibitem{Lindert:2017olm}
\hrefCMSnoop {}{J.~M. Lindert { et~al.}, ``Precise predictions for {\PV}+jets
  dark matter backgrounds'',} \textit{ Eur. Phys. J. C} \textbf{ 77} (2017)
  829,
  \href{http://dx.doi.org/10.1140/epjc/s10052-017-5389-1}{\doi{10.1140/epjc/s10052-017-5389-1}},
  \href{http://www.arXiv.org/abs/1705.04664}{\texttt{arXiv:1705.04664}}.

\bibitem{Frederix:2012ps}
\hrefCMSnoop {}{R.~Frederix and S.~Frixione, ``Merging meets matching in
  {MC@NLO}'',} \textit{ JHEP} \textbf{ 12} (2012) 061,
  \href{http://dx.doi.org/10.1007/JHEP12(2012)061}{\doi{10.1007/JHEP12(2012)061}},
  \href{http://www.arXiv.org/abs/1209.6215}{\texttt{arXiv:1209.6215}}.

\bibitem{Kulesza:2018tqz}
A.~Kulesza\hrefCMSnoop {}{ { et~al.}, ``Associated production of a top quark
  pair with a heavy electroweak gauge boson at {NLO}+{NNLL} accuracy'',}
  \textit{ Eur. Phys. J. C} \textbf{ 79} (2019) 249,
  \href{http://dx.doi.org/10.1140/epjc/s10052-019-6746-z}{\doi{10.1140/epjc/s10052-019-6746-z}},
  \href{http://www.arXiv.org/abs/1812.08622}{\texttt{arXiv:1812.08622}}.

\bibitem{powheg:singlet:st}
\hrefCMSnoop {}{S.~Alioli, P.~Nason, C.~Oleari, and E.~Re, ``{NLO} single-top
  production matched with shower in {POWHEG}: $s$- and $t$-channel
  contributions'',} \textit{ JHEP} \textbf{ 09} (2009) 111,
  \href{http://dx.doi.org/10.1088/1126-6708/2009/09/111}{\doi{10.1088/1126-6708/2009/09/111}},
  \href{http://www.arXiv.org/abs/0907.4076}{\texttt{arXiv:0907.4076}}.
  [Erratum: \DOI{10.1007/JHEP02(2010)011}].

\bibitem{powheg:singlet:tW}
\hrefCMSnoop {}{E.~Re, ``Single-top {Wt-channel} production matched with parton
  showers using the {POWHEG} method'',} \textit{ Eur. Phys. J. C} \textbf{ 71}
  (2011) 1547,
  \href{http://dx.doi.org/10.1140/epjc/s10052-011-1547-z}{\doi{10.1140/epjc/s10052-011-1547-z}},
  \href{http://www.arXiv.org/abs/1009.2450}{\texttt{arXiv:1009.2450}}.

\bibitem{bib:singletxs:t:1}
M.~Aliev\hrefCMSnoop {}{ { et~al.}, ``{HATHOR: HAdronic Top and Heavy quarks
  crOss section calculatoR}'',} \textit{ Comput. Phys. Commun.} \textbf{ 182}
  (2011) 1034,
  \href{http://dx.doi.org/10.1016/j.cpc.2010.12.040}{\doi{10.1016/j.cpc.2010.12.040}},
  \href{http://www.arXiv.org/abs/1007.1327}{\texttt{arXiv:1007.1327}}.

\bibitem{bib:singletxs:t:2}
P.~Kant\hrefCMSnoop {}{ { et~al.}, ``{HatHor} for single top-quark production:
  {Updated} predictions and uncertainty estimates for single top-quark
  production in hadronic collisions'',} \textit{ Comput. Phys. Commun.}
  \textbf{ 191} (2015) 74,
  \href{http://dx.doi.org/10.1016/j.cpc.2015.02.001}{\doi{10.1016/j.cpc.2015.02.001}},
  \href{http://www.arXiv.org/abs/1406.4403}{\texttt{arXiv:1406.4403}}.

\bibitem{bib:singletxs:tw:1}
\hrefCMSnoop {}{N.~Kidonakis, ``Two-loop soft anomalous dimensions for single
  top quark associated production with {\PWm} or {$\PH^{-}$}'',} \textit{ Phys.
  Rev. D} \textbf{ 82} (2010) 054018,
  \href{http://dx.doi.org/10.1103/PhysRevD.82.054018}{\doi{10.1103/PhysRevD.82.054018}},
  \href{http://www.arXiv.org/abs/1005.4451}{\texttt{arXiv:1005.4451}}.

\bibitem{Gehrmann:2014fva}
T.~Gehrmann\hrefCMSnoop {}{ { et~al.}, ``{$\PWp\PWm$} production at hadron
  colliders in next to next to leading order {QCD}'',} \textit{ Phys. Rev.
  Lett.} \textbf{ 113} (2014) 212001,
  \href{http://dx.doi.org/10.1103/PhysRevLett.113.212001}{\doi{10.1103/PhysRevLett.113.212001}},
  \href{http://www.arXiv.org/abs/1408.5243}{\texttt{arXiv:1408.5243}}.

\bibitem{bib:mcfm:diboson}
\hrefCMSnoop {}{J.~M. Campbell, R.~K. Ellis, and C.~Williams, ``Vector boson
  pair production at the {LHC}'',} \textit{ JHEP} \textbf{ 07} (2011) 018,
  \href{http://dx.doi.org/10.1007/JHEP07(2011)018}{\doi{10.1007/JHEP07(2011)018}},
  \href{http://www.arXiv.org/abs/1105.0020}{\texttt{arXiv:1105.0020}}.

\bibitem{CMS:2023vzh}
\hrefCMSnoop {}{{CMS Collaboration}, ``Measurement of simplified template cross
  sections of the {Higgs} boson produced in association with {\PW} or {\PZ}
  bosons in the {$\PH\to\bbbar$} decay channel in proton-proton collisions at
  {$\sqrt{s}=13\TeV$}'',} \textit{ Phys. Rev. D} \textbf{ 109} (2024) 092011,
  \href{http://dx.doi.org/10.1103/PhysRevD.109.092011}{\doi{10.1103/PhysRevD.109.092011}},
  \href{http://www.arXiv.org/abs/2312.07562}{\texttt{arXiv:2312.07562}}.

\bibitem{Conway:2011in}
\hrefCMSnoop {}{J.~S. Conway, ``Incorporating nuisance parameters in
  likelihoods for multisource spectra'',} in \textit{ {PHYSTAT} (2011):
  {Workshop} on statistical issues related to discovery claims in search
  experiments and unfolding}, p.~115.
\newblock 2011.
\newblock \href{http://www.arXiv.org/abs/1103.0354}{\texttt{arXiv:1103.0354}}.
\newblock
\href{http://dx.doi.org/10.5170/CERN-2011-006.115}{\doi{10.5170/CERN-2011-006.115}}.
%%CITATION = ARXIV:1103.0354;%%.

\bibitem{CMS:2024onh}
\hrefCMSnoop {}{{CMS Collaboration}, ``The {CMS} statistical analysis and
  combination tool: {\textsc{Combine}}'',} \textit{ Comput. Softw. Big Sci.}
  \textbf{ 8} (2024) 19,
  \href{http://dx.doi.org/10.1007/s41781-024-00121-4}{\doi{10.1007/s41781-024-00121-4}},
  \href{http://www.arXiv.org/abs/2404.06614}{\texttt{arXiv:2404.06614}}.

\bibitem{Junk:1999kv}
\hrefCMSnoop {}{T.~Junk, ``Confidence level computation for combining searches
  with small statistics'',} \textit{ Nucl. Instrum. Meth. A} \textbf{ 434}
  (1999) 435,
  \href{http://dx.doi.org/10.1016/S0168-9002(99)00498-2}{\doi{10.1016/S0168-9002(99)00498-2}},
  \href{http://www.arXiv.org/abs/hep-ex/9902006}{\texttt{arXiv:hep-ex/9902006}}.

\bibitem{Read:2002hq}
\hrefCMSnoop {}{A.~L. Read, ``Presentation of search results: {The} {\CLs}
  technique'',} \textit{ J. Phys. G} \textbf{ 28} (2002) 2693,
  \href{http://dx.doi.org/10.1088/0954-3899/28/10/313}{\doi{10.1088/0954-3899/28/10/313}}.

\bibitem{Cowan:2010js}
\hrefCMSnoop {}{G.~Cowan, K.~Cranmer, E.~Gross, and O.~Vitells, ``Asymptotic
  formulae for likelihood-based tests of new physics'',} \textit{ Eur. Phys. J.
  C} \textbf{ 71} (2011) 1554,
  \href{http://dx.doi.org/10.1140/epjc/s10052-011-1554-0}{\doi{10.1140/epjc/s10052-011-1554-0}},
  \href{http://www.arXiv.org/abs/1007.1727}{\texttt{arXiv:1007.1727}}.
  [Erratum: \DOI{10.1140/epjc/s10052-013-2501-z}].

\bibitem{Harlander:2016hcx}
\hrefCMSnoop {}{R.~V. Harlander, S.~Liebler, and H.~Mantler, ``{SusHi} {Bento}:
  {Beyond} {NNLO} and the heavy-top limit'',} \textit{ Comput. Phys. Commun.}
  \textbf{ 212} (2017) 239,
  \href{http://dx.doi.org/10.1016/j.cpc.2016.10.015}{\doi{10.1016/j.cpc.2016.10.015}},
  \href{http://www.arXiv.org/abs/1605.03190}{\texttt{arXiv:1605.03190}}.

\bibitem{Eriksson:2009ws}
\hrefCMSnoop {}{D.~Eriksson, J.~Rathsman, and O.~Stal, ``{2HDMC}:
  {Two-Higgs-Doublet Model Calculator} physics and manual'',} \textit{ Comput.
  Phys. Commun.} \textbf{ 181} (2010) 189,
  \href{http://dx.doi.org/10.1016/j.cpc.2009.09.011}{\doi{10.1016/j.cpc.2009.09.011}},
  \href{http://www.arXiv.org/abs/0902.0851}{\texttt{arXiv:0902.0851}}.

\bibitem{Bernon:2015qea}
J.~Bernon\hrefCMSnoop {}{ { et~al.}, ``Scrutinizing the alignment limit in
  two-{Higgs}-doublet models: {$m_{\PSh} = 125\GeV$}'',} \textit{ Phys. Rev. D}
  \textbf{ 92} (2015) 075004,
  \href{http://dx.doi.org/10.1103/PhysRevD.92.075004}{\doi{10.1103/PhysRevD.92.075004}},
  \href{http://www.arXiv.org/abs/1507.00933}{\texttt{arXiv:1507.00933}}.

\end{thebibliography}\endgroup
